\documentclass[conference]{IEEEtran}
\IEEEoverridecommandlockouts

\usepackage{amsmath,graphicx}
\usepackage[T1]{fontenc}
\usepackage{amsfonts}
\usepackage{amssymb}
\usepackage{bm}
\usepackage[noadjust]{cite}
\usepackage[nolist,printonlyused]{acronym}
\usepackage{algorithm}
\usepackage{algpseudocode}
\usepackage{mathtools}

\newcommand\eqbydef{\stackrel{\mathclap{\normalfont\mbox{\tiny def}}}{=}}
\newcommand\eqa{\stackrel{\mathclap{\normalfont\mbox{\tiny (a)}}}{=}}
\newcommand\eqb{\stackrel{\mathclap{\normalfont\mbox{\tiny (b)}}}{=}}
\newcommand\eqc{\stackrel{\mathclap{\normalfont\mbox{\tiny (c)}}}{=}}

\DeclareFontFamily{U}{mathx}{\hyphenchar\font45}
\DeclareFontShape{U}{mathx}{m}{n}{
	<5> <6> <7> <8> <9> <10>
	<10.95> <12> <14.4> <17.28> <20.74> <24.88>
	mathx10
}{}
\DeclareSymbolFont{mathx}{U}{mathx}{m}{n}
\DeclareFontSubstitution{U}{mathx}{m}{n}
\DeclareMathAccent{\widebar}{0}{mathx}{"73}

\usepackage{tikz}
\usepackage{pgfplots}
\pgfplotsset{compat=newest}
\usetikzlibrary{spy}
\def\figsize{0.24\textwidth}
\def\plotsize{0.24\textwidth}

\usepackage{times}

\usepackage{glossaries}
\newacronym{ris}{RIS}{Reconfigurable intelligent surface}

\title{Tensor-Based Channel Estimation and Reflection Design for RIS-Aided Millimeter-Wave MIMO Communication Systems}

\author{Sepideh Gherekhloo, Khaled Ardah, André L. F. de Almeida, and Martin Haardt
	\thanks{{The authors gratefully acknowledge the support of the German Research Foundation (DFG) under contracts no. HA 2239/14-1 and no. HA 2239/6-2 and the support of CAPES/PRINT (Grant no. 88887.311965/2018-00). The research of André L. F. de Almeida is partially supported by the CNPq (Grant no. 306616/2016-5).}}
	\thanks{S. Gherekhloo, K. Ardah, and M. Haardt are with the Communications Research Laboratory (CRL), TU Ilmenau, Ilmenau, Germany (e-mail: \{khaled.ardah,sepideh.gherekhloo, martin.haardt\}@tu-ilmenau.de). A. L. F. de Almeida is with the Wireless Telecom Research Group (GTEL), Federal University of Cear\'a, Fortaleza, Brazil (e-mail: andre@gtel.ufc.br).}
}

\def\ka#1{\textcolor{black}{#1}}
\def\kr#1{\textcolor{black}{#1}}

\begin{document}
%\ninept
%
\maketitle
\begin{abstract}
In this work, we consider both channel estimation and reflection \kr{coefficient} design problems in point-to-point reconfigurable intelligent surface (RIS)-aided millimeter-wave (mmWave) MIMO communication systems. First, we show that by exploiting the low-rank nature of mmWave MIMO channels, the received training signals can be written as a low-rank multi-way tensor admitting a canonical polyadic \kr{(CP)} decomposition. Utilizing such a structure, a tensor-based RIS channel estimation method (termed TenRICE) is proposed, wherein the tensor factor matrices are estimated using an alternating least squares method. Using TenRICE, the transmitter-to-RIS and the RIS-to-receiver channels are efficiently and separately estimated, up to a trivial scaling factor. After that, we formulate the beamforming and RIS reflection \kr{coefficient} design as a spectral efficiency maximization \kr{task}. Due to its non-convexity, we propose a heuristic non-iterative two-step method, where the RIS reflection vector is obtained in a closed form using a Frobenius-norm maximization (FroMax) strategy. Our numerical results show that TenRICE has \ka{a} superior performance, compared to benchmark methods, approaching the Cramér–Rao lower bound with a low training overhead. Moreover, we show that FroMax achieves a comparable performance to benchmark methods with a lower complexity. 

\end{abstract}
%\vspace{-5pt}

\vspace{-5pt}
\section{Introduction} \label{sec:intro}
%\vspace{-10pt}
\glspl{ris} have been proposed recently as {a} cost-effective {technology} for reconfiguring the propagation channels in wireless communication systems \cite{irs}. An RIS is a 2D surface equipped with a large number of tunable units that can be realized using, e.g., inexpensive antennas or metamaterials and controlled in real-time to influence the communication channels without generating its own signals. %Therefore, RISs are a promising software-deﬁned architecture for realizing the concept of smart radio environments. 
Among its many applications, an RIS can be utilized as a solution to the signal-blockage problem in millimeter-wave (mmWave)-based communications by providing alternative and tunable RIS-aided channels.

Recently, \gls{ris}-aided communications have attracted great attention, due to their potential of improving the efficiency of wireless mobile communications. RIS reflection design, in particular, have been extensively investigated under various setups and objectives, see \cite{RISCapacity,BF1,RISsec,IRSSINR} and reference therein. However, due to the non-convexity of \ka{the} involved problems, relaxations and alternating optimization techniques are commonly used to obtain a local\ka{ly} optimal solution. For example, the authors in \cite{RISCapacity} considered the capacity maximization and proposed an alternating optimization approach to find a locally optimal solution by iteratively optimizing the transmit covariance matrix or one of the RIS reflection coefficients with the others being fixed. However, such an alternating approach increases the computational complexity and becomes a limiting factor in practice, especially in a massive RIS setup.

The vast majority of the existing works assume perfect channel state information (CSI) at the transceivers, see \cite{RISCapacity,BF1,RISsec,IRSSINR} , which can never be obtained in practice. Recently, \gls{ris}-aided channel estimation (CE) methods have been proposed, e.g., {in} \cite{LSOnOff,MVDR,MMSE,RISUL}. These works, however, \ka{require that} the number of training subframes \kr{is}, at least, equal to the number of \gls{ris} reflection units to \ka{obtain} an accurate CSI estimate, which increases the training overhead and complexity. To overcome these issues, several approaches have been \ka{studied}, e.g., by exploiting the low-rank nature of mmWave channels and the \ka{multidimensional} (i.e., tensor) structure of the received signals. The former allows the CE to be formulated as a sparse-recovery problem and solved using compressed sensing (CS) tools \cite{CS,ardah_icassp2020,ardah_icassp19}, which are known to require \ka{a} few measurements to have \ka{an accurate} estimate, see \cite{ardah2020trice,CSGrid,MatrixCom}. In \cite{ardah2020trice}, by exploiting the low-rank nature of the mmWave channels, we have proposed the TRICE framework, which formulates the CE in RIS-aided mmWave MIMO systems as a two-stage multidimensional sparse-recovery problem.  On the other hand, tensor-based signal modeling and processing methods offer fundamental advantages over their bilinear (matrix) counterparts, since they have the ability to \ka{improve the identifiability of the parameters due to the powerful uniqueness properties of tensor decompositions \cite{Andre2014}.} In \cite{andre}, it is shown that the received signals in \gls{ris}-aided MIMO communication systems can be written as a 3-way tensor admitting \ka{a} canonical polyadic (CP) decomposition. However, the proposed method in \cite{andre} assumes sub-6 GHz systems and, thus, requires a large number of training subframes, similarly to \cite{LSOnOff,MVDR,MMSE,RISUL}.

In this paper, we extend our TRICE framework in \cite{ardah2020trice} and propose a CP \textbf{Ten}sor decomposition method for \textbf{RI}S-aided \textbf{CE} in mmWave MIMO systems, termed TenRICE, by jointly exploiting the tensor structure of \kr{the} received signals and the low-rank nature of mmWave channels. Using the TenRICE method, the transmitter-to-RIS and the RIS-to-receiver channels can be estimated \ka{separately}, up to a trivial scaling factor.
%In particular, we show that the received signals in P2P \gls{ris}-aided millimeter-wave (mmWave) MIMO systems, in which the transmitter (TX) and the receiver (RX) are equipped with multi-antenna uniform linear arrays (ULAs), can be written as a 4-way tensor admitting CP decomposition \cite{Kolda}. 
%Therefore, the CE turns into estimating the tensor factor matrices, where several powerful methods exist, e.g., in \cite{ROEMER20132722,Lathauwer}, including the alternating least squares (ALS) \cite{andre_tensors}. Given the estimated factor matrices, the channel parameters can be recovered using, e.g., a simple 1D correlation-based scheme or CS tools \cite{NOMP,ardah_icassp19}, with an automatic angle-pairing. 
After that, we formulate the beamforming and the RIS reflection \kr{coefficient} design as a spectral efficiency (SE) maximization problem. Due to its non-convexity, we propose a heuristic non-iterative two-step solution, where the RIS reflection vector is obtained, \kr{in contrast to} \cite{RISCapacity}, in a closed form using a \textbf{Fro}benius-norm \textbf{Max}imization (FroMax) strategy. Our numerical results show that TenRICE has \ka{a} superior performance, compared to \ka{the} TRICE framework, approaching the Cramér–Rao bound (CRB). Moreover, we show that FroMax achieves a comparable performance to benchmark methods with a lower complexity. 
%These advantages indicate the efficiency of the proposed methods, which makes them appealing to practical applications.

% Using computer simulations, we show that TenRICE achieves superior performance, comparing to benchmark methods, with low training overhead and low-complexity while approaching the Cramér–Rao lower bound (CRLB).   

\section{System Model}

\begin{figure}
	\centering
	\includegraphics[width=0.9\linewidth]{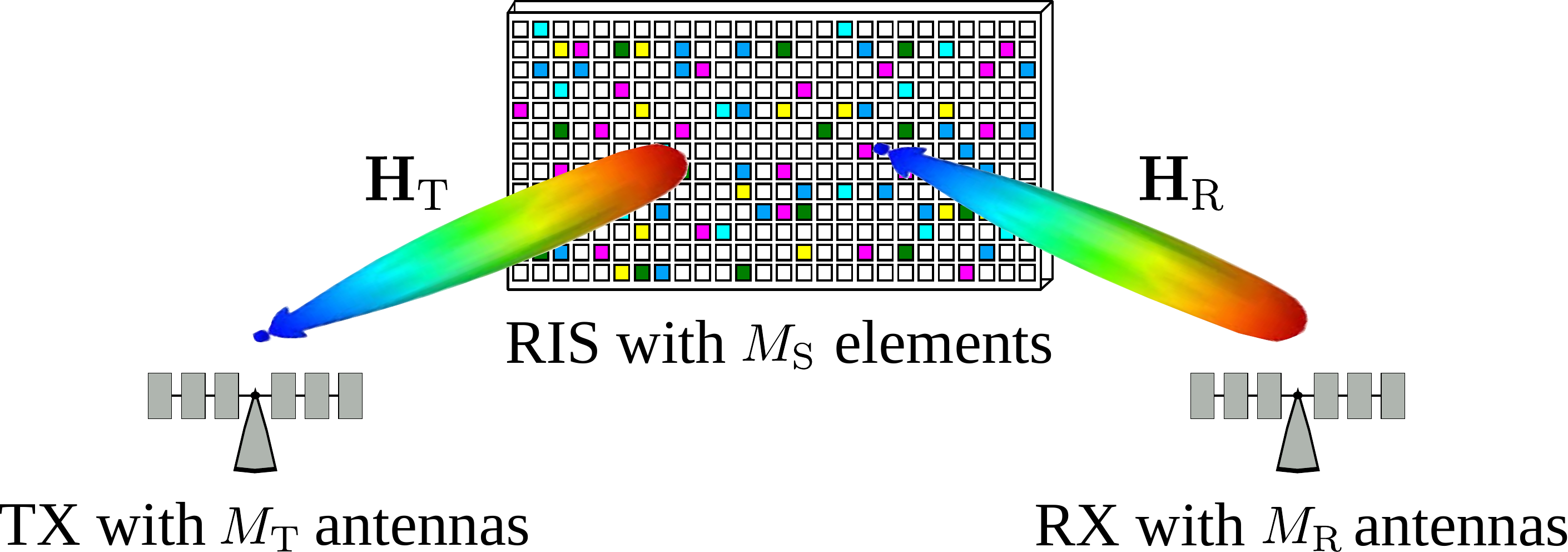}
	\caption{An \gls{ris}-aided MIMO mmWave communication system.}
	\label{fig:fig1}
%	\vspace{-15pt}
\end{figure}

In this paper\footnote{\textbf{Notation:} The transpose, the conjugate transpose (Hermitian), the Moore-Penrose pseudoinverse, the Kronecker product, and the Khatri-Rao product are denoted as ${\bm A}^{\mathsf{T}}$, ${\bm A}^{\mathsf{H}}$, ${\bm A}^{+}$, $\otimes$, and $\diamond$, respectively. Moreover, $\bm{1}_N$ is the all ones vector of length $N$, ${\bm I}_N$ is the $N\times N$ identity matrix, $\text{diag}\{{\bm a}\}$ forms a diagonal matrix ${\bm A}$ by putting the entries of the input vector ${\bm a}$ in its main diagonal, $\text{undiag}\{{\bm A}\}$ is the reverse of \ka{the} diag operator, $\text{vec}\{ {\bm A} \}$ forms a vector by staking the columns of ${\bm A}$ over each other, and the $n$-mode product of a tensor $\bm{\mathcal{A}}\in \mathbb{C}^{I_1\times I_2\times \dots,\times I_N}$ with a matrix ${\bm B}\in \mathbb{C}^{J\times I_n} $ is denoted as $\bm{\mathcal{A}} \times_n {\bm B}$. Throughout this paper, we assume that the \ka{singular} values of a given diagonal singular matrix are arranged in a decreasing order. Moreover, the following properties are used: \textit{Property~1}: $\text{vec}\{ {\bm A}{\bm B}{\bm C} \} = ({\bm C}^{\mathsf{T}}\otimes{\bm A}) \text{vec}\{{\bm B}\}$. \textit{Property~2}: ${\bm A}{\bm B}\diamond {\bm C}{\bm D} = ({\bm A}\otimes {\bm C})({\bm B}\diamond {\bm D})$. \textit{Property~3}: $({\bm A}\otimes {\bm C})({\bm B}\otimes {\bm D}) = {\bm A}{\bm B}\otimes{\bm C}{\bm D}$. \textit{Property~4:} Let ${\bm A}_1 \in \mathbb{C}^{J_1\times L_1}$ and ${\bm A}_2 \in \mathbb{C}^{J_2\times L_2}$. Then ${\bm A}_1 \otimes {\bm A}_2 = {\bm A}_1 \bm{\Omega}_1 \diamond {\bm A}_2 \bm{\Omega}_2$, where $\bm{\Omega}_{1} = {\bm I}_{L_1} \otimes \bm{1}^{\mathsf{T}}_{L_2}$ and $\bm{\Omega}_{2} = \bm{1}^{\mathsf{T}}_{L_1} \otimes {\bm I}_{L_2}$ so that  $\bm{\Omega}_{1} \diamond  \bm{\Omega}_{2} = {\bm I}_{L_1L_2}$. \kr{\textit{Property~5}: $\text{vec}\{ {\bm A}\text{diag}\{{\bm b}\}{\bm C} \} = ({\bm C}^{\mathsf{T}}\diamond{\bm A}) {\bm b}$}.}, we consider an \gls{ris}-aided mmWave MIMO communication system as depicted in Fig. \ref{fig:fig1}, where a transmitter (TX) with $M_{\text{T}}$ antennas is communicating with a receiver (RX) with $M_{\text{R}}$ antennas via an RIS-aided MIMO channel. The direct channel between the TX and the RX is assumed unavailable or too weak, e.g., due to blockage. The \gls{ris} has $M_{\text{S}}$ inexpensive reflecting elements arranged uniformly with half-wavelength inter-element spacing on a rectangular surface with $M^{\text{v}}_\text{S}$ vertical and $M^{\text{h}}_\text{S}$ horizontal elements such that $M_{\text{S}} = M^{\text{v}}_\text{S} \cdot M^{\text{h}}_\text{S}$.

\begin{figure}
	\centering
	\includegraphics[width=0.8\linewidth]{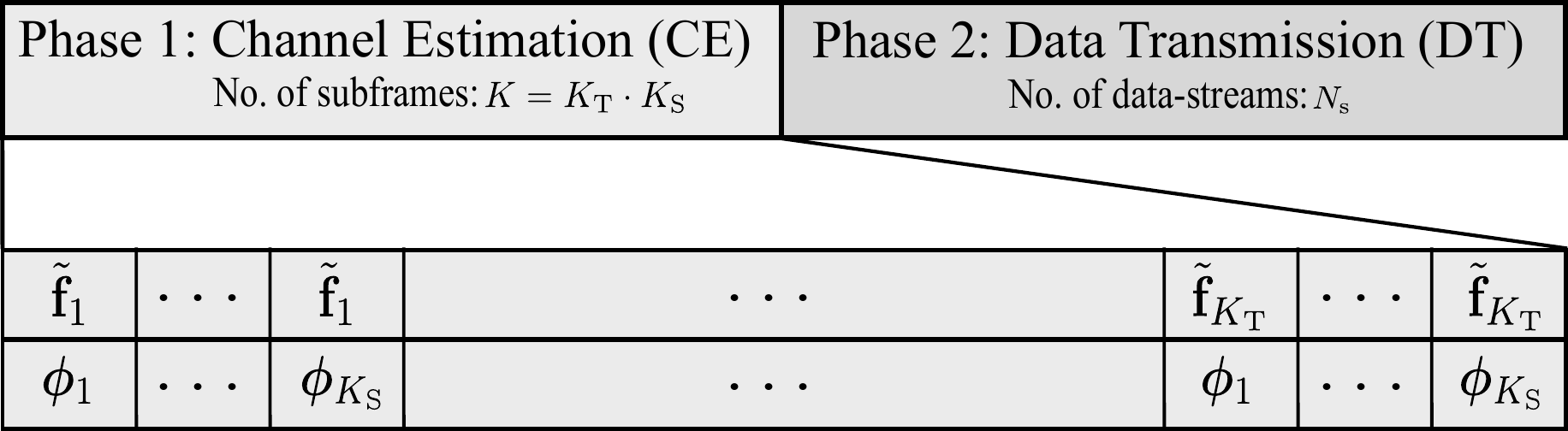}
	\caption{One channel coherence block.}
	\label{fig:fig2}
	\vspace{-15pt}
\end{figure}

Let ${\bm H}_{\text{T}} \in \mathbb{C}^{M_{\text{S}}\times M_{\text{T}}}$ be the TX to RIS channel and ${\bm H}_{\text{R}} \in \mathbb{C}^{M_{\text{R}}\times M_{\text{S}}}$ be the RIS to RX channel with $\mathbb{E}\{\Vert {\bm H}_{\text{T}} \Vert^2_{\text{F}}\} = M_{\text{S}} M_{\text{T}}$ and $\mathbb{E}\{\Vert {\bm H}_{\text{R}} \Vert^2_{\text{F}}\} = M_{\text{S}} M_{\text{R}}$. We assume a block-fading channel scenario, where ${\bm H}_{\text{T}}$ and ${\bm H}_{\text{R}}$ remain constant during every channel coherence block and change from block to block. We assume that every block is divided into two sub-blocks: one for CE and another for data transmission (DT), see Fig.~\ref{fig:fig2}. 

\textbf{In the CE phase}, we conduct a channel training procedure that occupies $K = K_{\text{T}}\cdot K_{\text{S}}$ subframes. The received signal at the RX at the $(s,t)$th subframe is given as
\begin{align}
	{\bm y}_{s,t} & = {{\bm W}^{\mathsf{H}}} {\bm H}_{\text{R}} \text{diag}\{\bm{\phi}_{s}\} {\bm H}_{\text{T}} \tilde{{\bm f}}_{t}s_t + {{\bm W}^{\mathsf{H}}} {\bm z}_{s,t} \in \mathbb{C}^{K_{\text{R}}},
\end{align}
where ${{\bm W}} \in \mathbb{C}^{M_{\text{R}}\times K_{\text{R}}}$ is a fixed training decoding matrix with $K_{\text{R}}$ beams, $\tilde{{\bm f}}_{t} \in \mathbb{C}^{M_{\text{T}}}$ is the $t$th training vector of \ka{the} TX with $\Vert \tilde{{\bm f}}_{t}\Vert^{2}_2 = 1$, $t \in \{1,\dots,K_\text{T}\}$, $\bm{\phi}_s \in \mathbb{C}^{M_{\text{S}}}$ is the $s$th training vector of \ka{the} RIS with $\big|[\bm{\phi}_s]_{[m]}\big| = \frac{1}{\sqrt{M_{\text{S}}}}, \forall m$, $s \in \{1,\dots,K_\text{S}\}$, $s_t \in \mathbb{C}$ is the unit-norm pilot symbol, and ${\bm z}_{s,t} \in \mathbb{C}^{M_{\text{R}}}$ is the additive white Gaussian noise vector {having zero-mean circularly symmetric complex-valued entries} with variance $\sigma^2$. We stack $\{{\bm y}_{s,t} \}_{t = 1}^{K_{\text{T}}}$ on top of each other as ${\bm y}_{s} =  [{\bm y}^{\mathsf{T}}_{s,1},\dots,{\bm y}^{\mathsf{T}}_{s,K_{\text{T}}}]^{\mathsf{T}}$ and after that we stack $\{{\bm y}_{s} \}_{s = 1}^{K_{\text{S}}}$ next to each other as ${\bm Y} = [{\bm y}_1,\dots, {\bm y}_{K_{\text{S}}}]$. Then, using Properties~2 and 5, the above measurement matrix ${\bm Y}$ can be written as
\begin{align}\label{eq1}
	{\bm Y} =  ({{\bm F}^{\mathsf{T}}} \otimes {{\bm W}^{\mathsf{H}}}) {\bm H}_{\text{c}}   {{\bm \Phi}} + {\bm Z} \in \mathbb{C}^{K_{\text{R}} K_{\text{T}} \times K_{\text{S}}},
\end{align}
where ${\bm H}_{\text{c}} = {\bm H}^{\mathsf{T}}_{\text{T}} \diamond {\bm H}_{\text{R}}$ represents the cascaded channel matrix, ${\bm Z} = [{\bm z}_1,\dots,{\bm z}_{K_{\text{S}}}]$, ${\bm z}_{s} =  [({{\bm W}^{\mathsf{H}}} {\bm z}_{s,1})^{\mathsf{T}},\dots,({{\bm W}^{\mathsf{H}}} {\bm z}_{s,K_{\text{T}}})^{\mathsf{T}} ]^{\mathsf{T}}$, ${{\bm F}} = [\tilde{{\bm f}}_1 s_1, \dots,\tilde{{\bm f}}_{K_{\text{T}}} s_{K_{\text{T}}}]$, and ${{\bm \Phi}} = [\bm{\phi}_1,\dots,\bm{\phi}_{K_{\text{S}}}]$. Given the measurement matrix ${\bm Y}$, the main goal of Section \ref{SectionCE} is to obtain an accurate estimate of ${\bm H}_{\text{T}}$ and ${\bm H}_{\text{R}}$, while keeping \ka{the} number of training subframes $K$ as small as possible.


\textbf{In the DT phase}, given the estimated channels $\widehat{{\bm H}}_{\text{R}}$ and $\widehat{{\bm H}}_{\text{T}}$, the TX first designs the precoding matrix ${{\bm P}} \in \mathbb{C}^{M_{\text{T}} \times N_\text{s}}$, the decoding matrix ${{\bm Q}} \in \mathbb{C}^{M_\text{R} \times N_\text{s}}$, and the RIS reflection \kr{coefficient} vector ${{\bm \omega}} \in \mathbb{C}^{M_{\text{S}}}$ with $\big|[\bm{\omega}]_{[m]}\big| = \frac{1}{\sqrt{M_{\text{S}}}}, \forall m$, to transmit \ka{the vector ${\bm s} \in \mathbb{C}^{N_\text{s}}$ of $N_\text{s}$ data \kr{streams}} with $\mathbb{E}[{\bm s} \bm{s}^{\mathsf{H}}] = {\bm I}_{N_\text{s}}$ to the RX. 
Therefore, the received signal vector at the RX is given as
\begin{align}\label{datasignal}
	{\bm y}  = {\bm Q}^{\mathsf{H}} {\bm H}_{\text{e}} {\bm P} {\bm s}  + {\bm Q}^{\mathsf{H}} {\bm z} \in \mathbb{C}^{N_\text{s}},
\end{align}    
%\ka{where $\bm{s} \in \mathbb{C}^{N_{\text{s}}}$ is the transmitted symbol vector, with $\mathbb{E}[{\bm s} \bm{s}^{\mathsf{H}}] = {\bm I}_{N_\text{s}}$.} 
where ${\bm H}_{\text{e}} = {\bm H}_{\text{R}} \text{diag}\{{\bm \omega} \} {\bm H}_{\text{T}}$ is the effective channel matrix. The system SE is given as 
\begin{align}\label{SE}
	\text{SE} = \log_2 \det ({\bm I}_{N_\text{s}} + {\bm R}^{-1}  {\bm Q}^{\mathsf{H}} {\bm H}_{\text{e}} {\bm P}  {\bm P}^{\mathsf{H}} {\bm H}^{\mathsf{H}}_{\text{e}}{\bm Q}  ),
\end{align}   
where ${\bm R} = \sigma^2{\bm Q}^{\mathsf{H}} {\bm Q}$ is the noise covariance matrix. In Section~\ref{SectionBeamforming}, we propose a non-iterative beamforming and RIS reflection \kr{coefficient} design method to maximize \ka{the} SE, where the RIS reflection vector is obtained in a closed form using a FroMax strategy.

\textbf{Channel model:} In mmWave-based communications \cite{Rappaport}, it was observed that the number of \ka{paths $L_\text{T}$ and $L_\text{R}$ for $\bm{H}_{\text{T}}$ and $\bm{H}_{\text{R}}$ respectively,} are very small compared to the number of antenna elements. This implies that $\text{rank}\{{\bm H}_{\text{T}}\} \leq L_\text{T}$ and $\text{rank}\{{\bm H}_{\text{R}}\} \leq L_\text{R}$. Therefore, similarly to \cite{ardah2020trice}, by assuming that the TX and the RX employ uniform linear arrays (ULAs)\footnote{The extension of the proposed methods to scenarios where {the} TX and/or {the} RX are {equipped} with uniform \kr{rectangular} arrays (URAs) is straightforward.}, ${\bm H}_{\text{T}}$ and ${\bm H}_{\text{R}}$ follow the geometric channel model, which \ka{can} be written as 
 \begingroup\makeatletter\def\f@size{9}\check@mathfonts
\def\maketag@@@#1{\hbox{\m@th\small\normalfont#1}}%  
\begin{equation} \label{channels}
\begin{aligned}
{\bm H}_{\text{T}} = & \frac{1}{\sqrt{L_{\text{T}}}} \sum_{\ell = 1}^{L_{\text{T}}} g_{\text{T},\ell} \bm{v}_{\text{2D}}(\mu_{{\text{T}},\ell}^{\text{v}},\mu_{{\text{T}},\ell}^{\text{h}}) \bm{v}_{\text{1D}}(\psi_{{\text{T}},\ell})^{\mathsf{T}} = & {\bm B}_{\text{T}} {\bm G}_{\text{T}}{\bm A}^{\mathsf{T}}_{\text{T}},\\ 
 {\bm H}_{\text{R}} = &  \frac{1}{\sqrt{L_{\text{R}}}} \sum_{\ell = 1}^{L_{\text{R}}} g_{\text{R},\ell} \bm{v}_{\text{1D}}(\psi_{{\text{R}},\ell}) \bm{v}_{\text{2D}}(\mu_{{\text{R}},\ell}^{\text{v}},\mu_{{\text{R}},\ell}^{\text{h}})^{\mathsf{T}} = & {\bm A}_{\text{R}}  {\bm G}_{\text{R}} {\bm B}^{\mathsf{T}}_{\text{R}},
\end{aligned}
\end{equation}\endgroup
where $g_{\text{X},\ell}\sim \mathcal{CN}(0,1)$ is the $\ell$th path gain, $\psi_{\text{T},\ell}\in [0,2\pi]$ is the $\ell$th \kr{direction-of-departure (DoD) spatial frequency from} the TX, $\psi_{\text{R},\ell} \in [0,2\pi] $ is the $\ell$th \kr{direction-of-arrival (DoA) spatial frequency} at the RX, $\mu^{\text{h}}_{\text{T},\ell} \in [0,2\pi]$ and $\mu^{\text{v}}_{\text{T},\ell} \in [0,\pi]$ are the $\ell$th \kr{horizontal and vertical DoA} \ka{spatial frequencies} at the RIS, while $\mu^{\text{h}}_{\text{R},\ell} \in [0,2\pi]$ and $ \mu^{\text{v}}_{\text{R},\ell}\in [0,\pi]$ are the $\ell$th \kr{horizontal and vertical DoD spatial frequencies from} the RIS. In (\ref{channels}), the 1D and the 2D array steering vectors are given as ${\bm v}_{{\text{1D}}}(\nu) = [1, e^{j \nu}, \dots,e^{j (M-1)  \nu}]^{\mathsf{T}} \in\mathbb{C}^{M}$ and $\bm{v}_{\text{2D}}(\nu^{\text{v}},\nu^{\text{h}}) = {\bm v}_{{\text{1D}}}(\nu^{\text{v}}) \diamond {\bm v}_{{\text{1D}}}(\nu^{\text{h}})$, respectively, where ${\bm v}_{{\text{1D}}}(\nu^{\text{v}}) \in\mathbb{C}^{M^{\text{v}}}$ and ${\bm v}_{{\text{1D}}}(\nu^{\text{h}}) \in\mathbb{C}^{M^{\text{h}}}$. Moreover, ${\bm H}_{\text{T}}$ and ${\bm H}_{\text{R}}$ are written in a compact form by letting
${\bm A}_{\text{X}} = [{\bm v}_{{\text{1D}}}({\psi}_{\text{X},1}),\dots,{\bm v}_{{\text{1D}}}({\psi}_{\text{X},L_{\text{X}}})] \in\mathbb{C}^{M_{\text{X}} \times L_{\text{X}} }$, ${\bm B}_{\text{X}} = {\bm B}^{\text{v}}_{\text{X}} \diamond {\bm B}^{\text{h}}_{\text{X}}$, ${\bm B}^{\text{Y}}_{\text{X}} = [{\bm v}_{{\text{1D}}}({\mu}^{\text{Y}}_{\text{X},1}),\dots,{\bm v}_{{\text{1D}}}({\mu}^{\text{Y}}_{\text{X},L_{\text{X}}})] \in\mathbb{C}^{M_{\text{S}}^{\text{Y}} \times L_{\text{X}} }$, and ${\bm G}_{\text{X}} = \frac{1}{\sqrt{L_{\text{X}}}}  \text{diag}\{ {g}_{\text{X},1},\dots, {g}_{\text{X},L_{\text{X}}}  \}$ for $\text{X} \in \{\text{T, R}\}$, $\text{Y}\in\{\text{v},\text{h}\}$.

 \section{Phase 1: The Proposed CE Method (TenRICE)}\label{SectionCE}
In this section, we propose our \textbf{Ten}sor-based \textbf{RI}S-aided \textbf{CE} (TenRICE) \kr{algorithm} by jointly exploiting the low-rank nature of mmWave channels and the tensor structure of received signals. By utilizing the channels model in (\ref{channels}), the cascaded channel matrix ${\bm H}_{\text{c}} = {\bm H}^{\text{T}}_{\text{T}} \diamond {\bm H}_{\text{R}}$ in (\ref{eq1}) can be written as  
 \begin{align}\label{h}
 	{\bm H}_{\text{c}} & =  ({\bm A}_{\text{T}} {\bm G}_{\text{T}}{\bm B}^{\mathsf{T}}_{\text{T}} \diamond {\bm A}_{\text{R}} {\bm G}_{\text{R}}{\bm B}^{\mathsf{T}}_{\text{R}}) 
 	\eqa  ({\bm A}_{\text{T}} \otimes {\bm A}_{\text{R}}) {\bm G} {\bm B},
 \end{align}
 where \ka{${\bm G} = {\bm G}_{\text{T}} \otimes {\bm G}_{\text{R}}  \in \mathbb{C}^{L \times L} $, ${\bm B} = {\bm B}^{\mathsf{T}}_{\text{T}} \diamond  {\bm B}^{\mathsf{T}}_{\text{R}}  \in \mathbb{C}^{L \times M_{\text{S}}} $}, $L = L_{\text{R}} \cdot L_{\text{T}}$, and $\eqa$ is obtained from Property~2. In \cite{ardah2020trice}, we have shown that ${{\bm B}}$ can be expressed as ${{\bm B}} = ({{\bm B}}_{\text{v}} \diamond  {{\bm B}}_{\text{h}})^{\mathsf{T}}$, where $ {{\bm B}}_{\text{v}} = [{\bm v}_{\text{1D} }({\mu}^{\text{v}}_{1} ), \dots, {\bm v}_{\text{1D} }({\mu}^{\text{v}}_{L})] \in \mathbb{C}^{M^{\text{v}}_\text{S} \times L}$, ${{\bm B}}_{\text{h}} = [{\bm v}_{\text{1D} }({\mu}^{\text{h}}_{1} ), \dots, {\bm v}_{\text{1D} }({\mu}^{\text{h}}_{L})]  \in \mathbb{C}^{M^{\text{h}}_\text{S} \times L}$, ${\mu}^{\text{v}}_{n} = {\mu}^{\text{v}}_{\text{T},\ell} + {\mu}^{\text{v}}_{\text{R},k}$, ${\mu}^{\text{h}}_{n} = {\mu}^{\text{h}}_{\text{T},\ell} + {\mu}^{\text{h}}_{\text{R},k}$, $\ell \in \{1,\dots, L_\text{T} \}$, $k \in \{1,\dots, L_\text{R} \}$, and $n = (\ell-1)\cdot L_\text{R} + k \in \{1,\dots, L\}$. Then, {using Property 2,} (\ref{h}) can be rewritten as 
  \begin{align}\label{h2}
 	{\bm H}_{\text{c}} =  ({\bm A}_{\text{T}} \otimes {\bm A}_{\text{R}}) {\bm G} ({{\bm B}}_{\text{v}} \diamond  {{\bm B}}_{\text{h}})^{\mathsf{T}},
 \end{align}
which is characterized by the following \ka{spatial frequency} vectors: $\bm{\psi}_{\text{R}} = [\psi_{\text{R},1}, \dots, \psi_{\text{R},L_{\text{R}}}]^{\mathsf{T}}$, $\bm{\psi}_{\text{T}} = [\psi_{\text{T},1}, \dots, \psi_{\text{T},L_{\text{T}}}]^{\mathsf{T}}$, $\bm{\mu}^{\text{h}} = [{\mu}^{\text{h}}_{1}, \dots, {\mu}^{\text{h}}_{L}  ]^{\mathsf{T}}$, and $\bm{\mu}^{\text{v}} = [{\mu}^{\text{v}}_{1}, \dots, {\mu}^{\text{v}}_{L}  ]^{\mathsf{T}}$ that define  ${{\bm  A}}_{\text{R}}$, ${{\bm  A}}_{\text{T}}$, ${\bm  B}_{\text{h}}$, and ${\bm  B}_{\text{v}}$, respectively. Therefore, to obtain an estimate of ${\bm H}_{\text{c}}$, it is sufficient to obtain an estimate of the above vectors from the measurement matrix ${\bm Y}$ in (\ref{eq1}), including the path gains vector ${\bm g} = \text{undiag}\{ {\bm G}\}$. In \cite{ardah2020trice}, we have proposed a two-stage framework, termed TRICE, which estimates $\bm{\psi}_{\text{R}}$ and $\bm{\psi}_{\text{T}}$ in the first stage \ka{as well as} $\bm{\mu}^{\text{h}}$, $\bm{\mu}^{\text{v}}$, and ${\bm g}$ in the second stage using any efficient multidimensional sparse-recovery technique, like CS \cite{ardah_icassp19} and ESPRIT \cite{StESBRIT}. To further improve the performance of \ka{the} TRICE framework, we propose in the following \ka{the} TenRICE method by exploiting the tensor structure of the measurement matrix ${\bm Y}$.

We assume that the RIS \kr{reflection coefficient matrix during the training phase} has a Kronecker structure given as $ {\bm \Phi} = {\bm \Phi}_{\text{v}} \otimes {\bm \Phi}_{\text{h}}$, where ${\bm \Phi}_{\text{v}} \in \mathbb{C}^{M^{\text{v}}_{\text{S}} \times K^{\text{v}}_{\text{S}}}$, ${\bm \Phi}_{\text{h}} \in \mathbb{C}^{M^{\text{h}}_{\text{S}} \times K^{\text{h}}_{\text{S}}}$, and $K^{\text{v}}_{\text{S}}\cdot K^{\text{h}}_{\text{S}} = K_{\text{S}}$. By substituting (\ref{h}) into (\ref{eq1}), the vectorized form of ${\bm Y}$, i.e., ${\bm y} = \text{vec}\{{\bm Y}\}$ can be written as
\begin{align}\label{eq3}
{\bm y} 
%&= \text{vec}\{  ({\bm F}^{\mathsf{T}} \otimes {\bm W}^{\mathsf{H}}) ({\bm A}_{\text{T}} \otimes {\bm A}_{\text{R}}) {\bm G} ({{\bm B}}_{\text{v}} \diamond  {{\bm B}}_{\text{h}})^{\mathsf{T}}  {\bm \Phi}  \}  +  {\bm z} \nonumber \\
%
& \eqa \text{vec}\{  ( {\bm F}^{\mathsf{T}} {\bm A}_{\text{T}} \otimes {\bm W}^{\mathsf{H}} {\bm A}_{\text{R}}) {\bm G} ({{\bm B}}_{\text{v}} \diamond  {{\bm B}}_{\text{h}})^{\mathsf{T}}   {\bm \Phi}  \}  +  {\bm z} \nonumber \\
& \eqb \text{vec}\{  ( {\bm F}^{\mathsf{T}}{{\bm A}}_{\text{T}}  \bm{\Omega}_\text{T} \diamond  {\bm W}^{\mathsf{H}}{{\bm A}}_{\text{R}}  \bm{\Omega}_\text{R} ) {\bm G} ({{\bm B}}_{\text{v}} \diamond  {{\bm B}}_{\text{h}})^{\mathsf{T}}   {\bm \Phi}  \}  +  {\bm z} \nonumber \\
& \eqc ({\bm \Phi}^{\mathsf{T}}_{\text{v}}{{\bm B}}_{\text{v}} \diamond {\bm \Phi}^{\mathsf{T}}_{\text{h}}{{\bm B}}_{\text{h}}  \diamond {\bm F}^{\mathsf{T}}{{\bm A}}_{\text{T}}  \bm{\Omega}_\text{T} \diamond  {\bm W}^{\mathsf{H}}{{\bm A}}_{\text{R}}  \bm{\Omega}_\text{R}) {\bm g}    + {\bm z},
\end{align} 
where ${\bm z} = \text{vec}\{{\bm Z}\}$ and ${\bm g} = \text{undiag}\{{\bm G}\}$. Moreover, $\eqa$, $\eqb$, and $\eqc$ are obtained by applying Properties 1,2, and~4, where $\bm{\Omega}_{\text{T}} \eqbydef {\bm I}_{L_{\text{T}}} \otimes \bm{1}^{\mathsf{T}}_{L_{\text{R}}}$ and $\bm{\Omega}_{\text{R}} \eqbydef \bm{1}^{\mathsf{T}}_{L_{\text{T}}} \otimes {\bm I}_{L_{\text{R}}}$. From (\ref{eq3}), we observe that ${\bm y}$ is the vectorized form of the transposed $4$-mode unfolding of a 4-way tensor $\bm{\mathcal{Y}} \in \mathbb{C}^{K_\text{R} \times K_\text{T} \times K^{\text{h}}_{\text{S}} \times K^{\text{v}}_{\text{S}} }$, i.e., ${\bm y} = [\bm{\mathcal{Y}}]^{\mathsf{T}}_{(4)}$ that admits a \ka{constrained} CP decomposition as \ka{\cite{Andre2008, Andre2014}}
 \begin{align}\label{tensor}
 	\bm{\mathcal{Y}} = \ka{{\bm{\mathcal{I}}}_{4,L}} \times_{1} {\bar{\bm{A}}}_{\text{R}} \bm{\Omega}_\text{R} \times_{2} {\bar{\bm{A}}}_{\text{T}} \bm{\Omega}_\text{T}  \times_{3} {\bar{\bm{B}}}_{\text{h}} \times_{4} {\bar{\bm{B}}}_{\text{v}} + \bm{\mathcal{Z}},
 \end{align}
 where $\bm{\mathcal{Z}}$ is the noise tensor, \ka{${\bm{\mathcal{I}}}_{4,L} \in \mathbb{C}^{L \times L \times L \times L}$ is a super-diagonal tensor with ones on the super diagonal}, and
 \begin{align}
 	{\bar{\bm{A}}}_{\text{R}} &=  {\bm W}^{\mathsf{H}}{{\bm A}}_{\text{R}} = {\bm W}^{\mathsf{H}} [{\bm v}_{\text{1D} }({\psi}_{\text{R},1} ), \dots, {\bm v}_{\text{1D} }({\psi}_{\text{R},L_{\text{R}} } ) ) ], \\ 
 	{\bar{\bm{A}}}_{\text{T}} &=  {\bm F}^{\mathsf{T}}{{\bm A}}_{\text{T}} = {\bm F}^{\mathsf{T}} [{\bm v}_{\text{1D} }({\psi}_{\text{T},1} ), \dots, {\bm v}_{\text{1D} }({\psi}_{\text{T},L_{\text{T}}} ) ) ], \\
 	{\bar{\bm{B}}}_{\text{h}} & = {\bm \Phi}^{\mathsf{T}}_{\text{h}}{{\bm B}}_{\text{h}} = {\bm \Phi}^{\mathsf{T}}_{\text{h}} [{\bm v}_{\text{1D} }({\mu}^{\text{h}}_{1} ), \dots, {\bm v}_{\text{1D} }({\mu}^{\text{h}}_{ L } ) ]  , \\
 	{\bar{\bm{B}}}_{\text{v}} & = {\bm \Phi}^{\mathsf{T}}_{\text{v}}{{\bm B}}_{\text{v}} {\bm G} = {\bm \Phi}^{\mathsf{T}}_{\text{v}} [{\bm v}_{\text{1D} }({\mu}^{\text{v}}_{1} ), \dots, {\bm v}_{\text{1D} }({\mu}^{\text{v}}_{ L } ) ]  {\bm G}.
 \end{align}

The \ka{$n$-mode unfoldings of tensor $\bm{\mathcal{Y}}$, for $n \in \{1, 2, 3, 4\}$} can be expressed as
% \footnote{Clearly, ${\bm y}$ in (\ref{eq3}) is the vectorized form of $[\bm{\mathcal{Y}}]^{\mathsf{T}}_{(4)}$, which can be readily obtained by applying Property 1 and Property 2.} 
 \begin{align}
 	[\bm{\mathcal{Y}}]_{(1)} & = {\bar{\bm{A}}}_{\text{R}} \bm{\Omega_{\text{R}}} ({\bar{\bm{B}}}_{\text{v}} \diamond {\bar{\bm{B}}}_{\text{h}} \diamond {\bar{\bm{A}}}_{\text{T}} \bm{\Omega}_\text{T})^{\mathsf{T}} + [\bm{\mathcal{Z}}]_{(1)}\\
 	[\bm{\mathcal{Y}}]_{(2)} &= {\bar{\bm{A}}}_{\text{T}} \bm{\Omega_{\text{T}}} ({\bar{\bm{B}}}_{\text{v}} \diamond {\bar{\bm{B}}}_{\text{h}} \diamond {\bar{\bm{A}}}_{\text{R}} \bm{\Omega_{\text{R}}} )^{\mathsf{T}} + [\bm{\mathcal{Z}}]_{(2)}\\
 	[\bm{\mathcal{Y}}]_{(3)} &= {\bar{\bm{B}}}_{\text{h}}  ({\bar{\bm{B}}}_{\text{v}}  \diamond{{\bar{\bm{A}}}}_{\text{T}} \bm{\Omega_{\text{T}}}   \diamond {{\bar{\bm{A}}}}_{\text{R}} \bm{\Omega_{\text{R}}}  )^{\mathsf{T}} + [\bm{\mathcal{Z}}]_{(3)}  \\
 	[\bm{\mathcal{Y}}]_{(4)} &= {\bar{\bm{B}}}_{\text{v}}  ({\bar{\bm{B}}}_{\text{h}}  \diamond{{\bar{\bm{A}}}}_{\text{T}} \bm{\Omega_{\text{T}}}   \diamond {\bar{\bm{A}}}_{\text{R}} \bm{\Omega_{\text{R}}}  )^{\mathsf{T}} + [\bm{\mathcal{Z}}]_{(4)} .
 \end{align}
 
 Given \kr{the measurement tensor} $\bm{\mathcal{Y}}$, the CE \kr{task} boils down to first estimating the tensor factor matrices.  
 Several techniques have been proposed to achieve this end, e.g., in \cite{Lathauwer,ardah2019wsa,ROEMER20132722}.  
 One of these techniques is the alternating least squares (ALS) \cite{andre_tensors}, which minimizes the data fitting error with respect to one of the factor matrices, with the other three being fixed. For example, to estimate ${\bar{\bm{A}}}_{\text{R}}$, assuming that ${\bar{\bm{A}}}_{\text{T}}$, ${\bar{\bm{B}}}_{\text{h}}$, and ${\bar{\bm{B}}}_{\text{v}}$ are fixed, the problem can be formulated as 
 \begingroup\makeatletter\def\f@size{9.8}\check@mathfonts
 \def\maketag@@@#1{\hbox{\m@th\small\normalfont#1}}%  
 \begin{align}\label{Arest}
 	{\bar{\bm{A}}}_{\text{R}} &= \underset{{\bar{\bm{A}}}_{\text{R}}}{\arg\min} \Big\Vert  [\bm{\mathcal{Y}}]_{(1)} - {\bar{\bm{A}}}_{\text{R}} \bm{\Omega_{\text{R}}} ({\bar{\bm{B}}}_{\text{v}} \diamond {\bar{\bm{B}}}_{\text{h}} \diamond {\bar{\bm{A}}}_{\text{T}} \bm{\Omega_{\text{T}}} )^{\mathsf{T}}   \Big\Vert^{2}_{\text{F}},
 \end{align}\endgroup
 which is a convex problem and can be solved using the LS method. Using the same methodology, ${\bar{\bm{A}}}_{\text{T}}$, ${\bar{\bm{B}}}_{\text{h}}$, and ${\bar{\bm{B}}}_{\text{v}}$ can be estimated similarly to (\ref{Arest}). Therefore, an ALS-based method can be used to estimate the four factor matrices as summarized in Algorithm \ref{als} (from step \ref{stepone} to step \ref{steplast}), which is guaranteed to converge monotonically to a local optimum point \cite{andre_tensors}.

 Let ${\hat{\bar{\bm{A}}}}_{\text{R}}$, ${\hat{\bar{\bm{A}}}}_{\text{T}}$, ${\hat{\bar{\bm{B}}}}_{\text{h}}$, and ${\hat{\bar{\bm{B}}}}_{\text{v}}$ denote the estimated factor matrices at the convergence of the iterative steps of Algorithm~\ref{als}. Then, the parameters \ka{associated} with each factor matrix can be recovered, e.g., \ka{via} a simple correlation-based scheme. For example, the $k$th entry of ${\bm{\psi}}_{\text{R}}$, i.e.,  $\psi_{\text{R},k}$ associated with the $k$th column vector of ${\hat{\bar{\bm{A}}}}_{\text{R}}$, i.e., ${\hat{\bar{\bm{a}}}}_{\text{R},k}$ can be recovered as
 \begin{align}\label{corelation}
 	\hat{\psi}_{\text{R},k} = \underset{\psi \in [0,2\pi]}{\arg\max} \frac{| {\hat{\bar{\bm{a}}}}^{\mathsf{H}}_{\text{R},k}  {\bm W}^{\mathsf{H}}  {\bm v}_{{\text{1D}}}({\psi})     |}{  \Vert {\hat{\bar{\bm{a}}}}_{\text{R},k}  \Vert \Vert {\bm W}^{\mathsf{H}}  {\bm v}_{{\text{1D}}}({\psi}) \Vert   }, 
 \end{align} 
 which can be efficiently implemented by first employing a coarse grid and then gradually refining it around the \kr{maximizing} grid points. Alternatively, (\ref{corelation}) can be interpreted as an off-grid sparse recovery problem, where efficient methods like, Newtonized OMP (NOMP) \cite{NOMP} can be readily applied to recover $\hat{\psi}_{\text{R},k}$ with high accuracy and low complexity. A similar approach can be used to recover the vectors $\bm{\psi}_{\text{T}}$, $\bm{\mu}^{\text{h}}$, and $\bm{\mu}^{\text{v}}$ from ${\hat{\bar{\bm{A}}}}_{\text{T}}$, ${\hat{\bar{\bm{B}}}}_{\text{h}}$, and ${\hat{\bar{\bm{B}}}}_{\text{v}}$, respectively.

 Next, using the estimated vectors $\hat{\bm{\psi}}_{\text{R}}$, $\hat{\bm{\psi}}_{\text{T}}$, $\hat{\bm{\mu}}^{\text{h}}$, and $\hat{\bm{\mu}}^{\text{v}}$ in step \ref{steprec}, we reconstruct ${\bm {\hat{A}}}_{\text{T}}$, ${\bm {\hat{A}}}_{\text{T}}$, ${\bm {\hat{B}}}_{\text{h}}$, and ${\bm {\hat{B}}}_{\text{v}}$. Then, the path gain vector ${{\bm g}}$ can be estimated from (\ref{eq3}) (or $[\bm{\mathcal{Y}}]^{\mathsf{T}}_{(4)}$) using a LS method as shown by step \ref{gest}. 
 Finally, the cascaded channel matrix $\widehat{{\bm H}}_{\text{c}}$ can be reconstructed as in step \ref{Hce}, which can be used to estimate $\widehat{{{\bm H}}}_{\text{T}} $ and $\widehat{{{\bm H}}}_{\text{R}} $, {up to trivial scaling factors}, using the LS Khatri-Rao factorization (LSKRF) method \cite{andre}.

 \begin{algorithm}[t]
 	\caption{\textbf{Ten}sor-based \textbf{RI}S-aided \textbf{CE} (TenRICE)}
 	\label{als}
 	\begingroup\makeatletter\def\f@size{9}\check@mathfonts
 	\def\maketag@@@#1{\hbox{\m@th\small\normalfont#1}}% 
 	\begin{algorithmic}[1]
 		\State{Input: Measurement tensor $\bm{\mathcal{Y}} \in \mathbb{C}^{K_\text{R} \times K_\text{T} \times K^{\text{h}}_{\text{S}} \times K^{\text{v}}_{\text{S}} }$ and $I_{\max}$}
 		\State{Output: Estimated channels $\widehat{{\bm H}}_{\text{T}} $ and $ \widehat{{\bm H}}_{\text{R}}$}
 		\State{Initialization: $\hat{\bar{\bm{B}}}^{(0)}_{\text{v}}$, ${\hat{\bar{\bm{B}}}}^{(0)}_{\text{h}}$, and ${\hat{\bar{\bm{A}}}}^{(0)}_{\text{T}}$, e.g., randomly }\label{stepone}
 	%	\State{Select $I_{\max}$ and set $i = 1$  }
 		%\Statex{ \textit{ ALS-based method for tensor factor matrices estimation} }
 		\While{not converged or $i< I_{\max}$}
 		\State{  ${\hat{\bar{\bm{A}}}}^{(i)}_{\text{R}} =  [\bm{\mathcal{Y}}]_{(1)} \Big[  \bm{\Omega_{\text{R}}} ({\hat{\bar{\bm{B}}}}^{(i-1)}_{\text{v}} \diamond {\hat{\bar{\bm{B}}}}^{(i-1)}_{\text{h}} \diamond {\hat{\bar{\bm{A}}}}^{(i-1)}_{\text{T}} \bm{\Omega_{\text{T}}})^{\mathsf{T}}   \Big]^{+}  $ }
 		
 		\State{  ${\hat{\bar{\bm{A}}}}^{(i)}_{\text{T}} =  [\bm{\mathcal{Y}}]_{(2)} \Big[  \bm{\Omega_{\text{T}}} ({\hat{\bar{\bm{B}}}}^{(i-1)}_{\text{v}} \diamond {\hat{\bar{\bm{B}}}}^{(i-1)}_{\text{h}} \diamond {\hat{\bar{\bm{A}}}}^{(i)}_{\text{R}} \bm{\Omega_{\text{R}}})^{\mathsf{T}}   \Big]^{+}  $ }
 		
 		\State{  ${\hat{\bar{\bm{B}}}}^{(i)}_{\text{h}} =  [\bm{\mathcal{Y}}]_{(3)} \Big[  ({\hat{\bar{\bm{B}}}}^{(i-1)}_{\text{v}} \diamond {\hat{\bar{\bm{A}}}}^{(i)}_{\text{T}} \bm{\Omega_{\text{T}}} \diamond {\hat{\bar{\bm{A}}}}^{(i)}_{\text{R}} \bm{\Omega_{\text{R}}})^{\mathsf{T}}   \Big]^{+}  $ }
 		
 		\State{  ${\hat{\bar{\bm{B}}}}^{(i)}_{\text{v}} =  [\bm{\mathcal{Y}}]_{(4)} \Big[  ({\hat{\bar{\bm{B}}}}^{(i)}_{\text{h}} \diamond {\hat{\bar{\bm{A}}}}^{(i)}_{\text{T}} \bm{\Omega_{\text{T}}} \diamond {\hat{\bar{\bm{A}}}}^{(i)}_{\text{R}} \bm{\Omega_{\text{R}}})^{\mathsf{T}}   \Big]^{+}  $ }
 		\EndWhile\label{steplast}
 		\State{Recover $\hat{\bm{\psi}}_{\text{R}}$, $\hat{\bm{\psi}}_{\text{T}}$, $\hat{\bm{\mu}}^{\text{h}}$, $\hat{\bm{\mu}}^{\text{v}}$ using, e.g., (\ref{corelation}) or NOMP \cite{NOMP} } \label{steprec}
 		\State{Compute ${{\hat{\bm{g}}}} = \big[ {\bm \Phi}^{\mathsf{T}}_{\text{v}}{\hat{{\bm{B}}}}_{\text{v}} \diamond {\bm \Phi}^{\mathsf{T}}_{\text{h}}{\hat{{\bm{B}}}}_{\text{h}}  \diamond {\bm F}^{\mathsf{T}}{\hat{{\bm{A}}}}_{\text{T}}  \bm{\Omega}_\text{T} \diamond  {\bm W}^{\mathsf{H}}{\hat{{\bm{A}}}}_{\text{R}}  \bm{\Omega}_\text{R} \big]^{+}{\bm y} $} \label{gest}
 		\State{Reconstruct $\widehat{{\bm H}}_{\text{c}} = ({\hat{{\bm{A}}}}_{\text{R}} \otimes {\hat{{\bm{A}}}}_{\text{T}} ) \text{diag}\{{\hat{{\bm{g}}}}\} ({\hat{{\bm{B}}}}_{\text{v}} \diamond {\hat{{\bm{B}}}}_{\text{h}} )^{\mathsf{T}}  $ }\label{Hce}
 		\State{Estimate $\widehat{{\bm H}}_{\text{T}} $ and $ \widehat{{\bm H}}_{\text{R}}$ from $\widehat{{\bm H}}$ using \cite[Algorithm 1]{andre} }
 		
 	\end{algorithmic}\endgroup
 	
 \end{algorithm}

 \textbf{Uniqueness and identifiability conditions:} It is well known that the CP decomposition is unique up to scaling and permutation ambiguities under mild conditions \cite{LowRankTen,andre_tensors,Kolda,CCPD,Andre2010}. In general, the uniqueness of a CP decomposition is guaranteed by Kruskal's condition \cite{Kolda}, which is also known as the $k$-rank. However, due to the definitions of $\bm{\Omega}_{\text{R}}$ and $\bm{\Omega}_{\text{T}}$, the first two factor matrices, i.e.,  ${{\bar{\bm{A}}}}_{\text{R}} \bm{\Omega_{\text{R}}} = {\mathring{\bm{A}} }_{\text{R}}$ and ${{\bar{\bm{A}}}}_{\text{T}} \bm{\Omega_{\text{T}}} = {\mathring{\bm{A}}}_{\text{T}}$ contain repeated columns, where every column of ${\mathring{\bm{A}}}_{\text{R}}$ is repeated $L_\text{T}$ times and every column of ${\mathring{\bm{A}}}_{\text{T}}$ is repeated $L_\text{R}$ times. This implies that the $k$-rank of ${\mathring{\bm{A}}}_{\text{R}}$ and ${\mathring{\bm{A}}}_{\text{T}}$ is equal to one. Therefore, the sufficient condition of \cite{Kolda} fails \cite{Andre2010}. 
 %\ka{For more details on the uniqueness properties of the constrained CP model in (\ref{corelation}), please refer to \cite{Andre2010}.} 
 As for Algorithm \ref{als}, which is an ALS-based algorithm, the identifiability in the LS sense requires that each of the following matrices: ${\bm C}_{\text{R}} = \bm{\Omega_{\text{R}}} ({\bm {\bar{B}}}_{\text{v}} \diamond {\bm {\bar{B}}}_{\text{h}} \diamond {\bm {\bar{A}}}_{\text{T}} \bm{\Omega_{\text{T}}})^{\mathsf{T}} \in \mathbb{C}^{L_\text{R} \times J_{\text{R}} }$, 
 ${\bm C}_{\text{T}} = \bm{\Omega_{\text{T}}} ({\bm {\bar{B}}}_{\text{v}} \diamond {\bm {\bar{B}}}_{\text{h}} \diamond {\bm {\bar{A}}}_{\text{R}} \bm{\Omega_{\text{R}}})^{\mathsf{T}}  \in \mathbb{C}^{L_\text{T} \times J_{\text{T}} }$, 
 ${\bm C}_{\text{h}} = ({\bm {\bar{B}}}_{\text{v}} \diamond {\bm {\bar{A}}}_{\text{T}} \bm{\Omega_{\text{T}}} \diamond {\bm {\bar{A}}}_{\text{R}} \bm{\Omega_{\text{R}}})^{\mathsf{T}} \in \mathbb{C}^{L \times J^{\text{h}}_{\text{S}} }$, and 
 ${\bm C}_{\text{v}} = ({\bm {\bar{B}}}_{\text{h}} \diamond {\bm {\bar{A}}}_{\text{T}} \bm{\Omega_{\text{T}}} \diamond {\bm {\bar{A}}}_{\text{R}} \bm{\Omega_{\text{R}}})^{\mathsf{T}} \in \mathbb{C}^{L \times J^{\text{v}}_{\text{S}} }$ 
 to have a unique right Moore-Penrose \kr{pseudo-inverse}, i.e., full row-rank, where $J_{\text{R}} = K_\text{T}  K_{\text{S}}$, $J_{\text{T}} = K_\text{R}  K_{\text{S}}$, $J^{\text{h}}_{\text{S}}  = K_\text{R} K_\text{T}  K^{\text{v}}_{\text{S}}$, and $J^{\text{v}}_{\text{S}}  = K_\text{R} K_\text{T}  K^{\text{h}}_{\text{S}}$. 
 This requires that $J_{\text{R}} \geq L_\text{R}$, $J_{\text{T}} \geq L_\text{T}$, $J^{\text{h}}_{\text{S}} \geq L$, and $J^{\text{v}}_{\text{S}} \geq L$, where $L = L_\text{R} \cdot L_\text{T}$. Since $L_\text{R}$ and $L_\text{T}$ are practically very small (i.e., $\max\{L_\text{R},L_\text{T}\} \approx 3$ \cite{Rappaport}), the above conditions are easily satisfied. For example, assuming that the TX is in line-of-sight with the RIS, we have that $L_\text{T} = 1$, as it has been assumed in \cite{MMSE}.

 \textbf{Complexity analysis}: Assuming that the complexity of calculating the Moore-Penrose \kr{pseudo-inverse} of a $n\times m$ matrix is \ka{on the} order of $\mathcal{O}(\min\{n,m\}^3)$. Then, the complexity of the ALS steps in Alg. \ref{als} is \ka{on the} order of $\mathcal{O}\big( I_{\max} (L_{\text{R}}^3 + L_{\text{T}}^3 + 2L^3)  \big)$. Moreover, assuming that the \ka{NOMP} method from \cite{NOMP} is used in step~\ref{steprec}, then the complexity of recovering the channel parameters is \ka{on the} order of $\bar{L}( L_{\text{R}} + L_{\text{T}} + 2L)  )  \big)$, where $\bar{L}$ denotes the number of grid points used by NOMP in the sparse-coding stage. In comparison, the complexity of TRICE-CS \cite{ardah2020trice} is \ka{on the} order of $\mathcal{O}(L(K_\text{R} K_\text{T}(\bar{L}^2 + L + L^2)) + 2L^3 + LK_\text{S} \bar{L}^2) $ and \kr{the} Joint-CS method \cite{CSGrid} is \ka{on the} order of $\mathcal{O}(L(N_\text{R} K_\text{T} K_\text{S}(\bar{L}^4 + L + L^2)) + L^3)$. Clearly, TenRICE has a much lower complexity compared to both methods. The main reason is that TRICE and Joint-CS require multidimensional (xD) dictionaries (2D for TRICE and 4D for Joint-CS) compared to the 1D dictionary required by TenRICE. Moreover, in contrast to the TenRICE, TRICE and Joint-CS methods require a dictionary orthogonalization operation during the \ka{parameter} recovery \cite{OMPComp}, which is very complex especially with large dictionaries. %Such \ka{a} complex operation is not required by TenRICE thanks to its ALS stage, which decouples the recovery process between the channel parameters as shown by (\ref{corelation}).  
 
 %For example, let $L_{\text{R}} = L_{\text{T}} = 2$, $K_{\text{R}} = K_{\text{T}} = K_{\text{S}} = 8$, and ${\bar L} = 128$. Then, 4416 15936 
 
% 100 × (2^3 + 2^3 + 2 × 4^3) + 128 × (2 + 2 + 2 × 4)
 
\vspace{-5pt}

\section{Phase 2: The Proposed RIS Reflection Design Method (FroMax)}\label{SectionBeamforming}
In this section, given the estimated channels $\widehat{{\bm H}}_{\text{R}}$ and $\widehat{{\bm H}}_{\text{T}}$, we design the TX and the RX beamforming matrices and the RIS reflection \kr{coefficient} vector as a solution to the following SE maximization problem:  
\begin{equation}\label{Opt}
	\begin{aligned}
		\max_{{\bm Q}, {\bm P}, {\bm \omega} } \quad & \log_2 \det ({\bm I}_{N_\text{s}} + {\bm R}^{-1}  {\bm Q}^{\mathsf{H}} \widehat{{\bm H}}_{\text{e}} {\bm P} {\bm P}^{\mathsf{H}} \widehat{{\bm H}}^{\mathsf{H}}_{\text{e}} {\bm Q}  ) \\
		\text{s.t.} \quad &\Vert {\bm P} \Vert_{\text{F}}^{2} \leq \ka{P}_{\max} \text{ and }
 \big|[{\bm \omega}]_{[m]}\big| = {1}/{\sqrt{M_{\text{S}}}} , \forall m,
	\end{aligned}
\end{equation}
where $\widehat{{\bm H}}_{\text{e}} \eqbydef \widehat{{\bm H}}_{\text{R}} \text{diag}\{{\bm \omega}\} \widehat{{\bm H}}_{\text{T}}$ and $\ka{P}_{\max}$ is the transmit power at the TX. Note that (\ref{Opt}) is non-convex, since the objective function is non-concave over $\bm{\omega}$ and the constant modulus constraints are non-convex functions. Moreover, ${\bm P}$, ${\bm Q}$, and ${\bm \omega}$ \kr{depend on each other}, which makes (\ref{Opt}) a difficult problem to solve. 
%In \cite{RISCapacity}, by decoupling the optimization procedure between ${\bm W}$ and ${\bm F}$, on the on hand, and $\bm{\phi}$, on the other hand, an alternating optimization algorithm is proposed to solve problem (\ref{Opt}), in which the RIS reflection elements $e^{j\phi_m}, \forall m$, are updated sequentially, one at a time, with the others being fixed. Consequently, the computational complexity becomes a limiting factor in practice, especially in a massive RIS setup. 
In the following, we propose a non-iterative solution to (\ref{Opt}), which has a comparable performance to that of \cite{RISCapacity}, but with a much lower complexity.

Initially, it is not hard to see  that for any given $\bm{\omega}$, (\ref{Opt}) reduces to a single-user multi-stream MIMO communication system. Let $\widehat{{\bm H}}_{\text{e}} = {\bm U}_{\widehat{{\bm H}}_{\text{e}}} \bm{\Sigma}_{\widehat{{\bm H}}_{\text{e}}} {\bm V}^{\mathsf{H}}_{\widehat{{\bm H}}_{\text{e}}}$ be the singular value decomposition (SVD) of $\widehat{{\bm H}}_{\text{e}}$. Then, the optimal fully-digital\footnote{Here, we note that in mmWave-based communications, hybrid analog-digital (HAD) beamforming architectures \cite{heath_overview,sepideh,ardah_unify,ardah_tvt} are generally assumed to reduce the power consumption. However, since in this section we focus \kr{on} the RIS reflection \kr{coefficient} design, we assume fully-digital beamforming architectures at \ka{the} TX and \ka{the} RX, to simplify the exposition.} solutions to ${\bm Q}$ and ${\bm P}$, for fixed ${\bm \omega}$, are given as 
\begin{align}\label{WF}
{\bm Q} = {\bm U}_s \text{ and } {\bm P} = {\bm V}_s \text{diag}\{\sqrt{p_1},\dots,\sqrt{p_{N_\text{s}}}\},
\end{align}   
where ${\bm U}_\text{s} = [{\bm U}_{\widehat{{\bm H}}_{\text{e}}}]_{[:,1:N_\text{s}]}$, ${\bm V}_\text{s} = [{\bm V}_{\widehat{{\bm H}}_{\text{e}}}]_{[:,1:N_\text{s}]}$, and $\{p_i\}_{i = 1}^{N_\text{s}}$ are the power allocations found using the waterfilling method \cite{Palomar2005} such that $\sum_{i = 1}^{N_\text{s}} p_i = \ka{P}_{\max}$. Consequently, ${\bm Q}^{\mathsf{H}} {\bm Q}={\bm I}_{N_\text{s}}$, $\bm{\Sigma}_s = {\bm U}^{\mathsf{H}}_s \widehat{{\bm H}}_{\text{R}} \text{diag}\{{\bm \omega}\}  \widehat{{\bm H}}_{\text{T}} {\bm V}_s  = \text{diag}\{\alpha_{1}, \dots, \alpha_{N_\text{s}}\}$, and the SE expression in (\ref{SE}) simplifies to 
\begin{align}\label{SE2}
	\text{SE} =  \sum_{i = 1}^{N_\text{s}} \log_2(1 + \frac{1}{\sigma^2} \alpha^2_{i} p_i ),
\end{align} 
where $\alpha_{i}$ is the $i$th dominant singular value in $\bm{\Sigma}_{\widehat{{\bm H}}_{\text{e}}}$. In the following, we turn our attention to \ka{the} RIS reflection \kr{coefficient} design and propose an efficient non-iterative solution to \kr{find} ${\bm \omega}$ based on a FroMax design strategy.  

\textbf{FroMax-1}: As a baseline method, the RIS reflection vector is found as a solution to   
\begin{equation}\label{PhiP2}
	\begin{aligned}
		\bm{\omega} = & \text{ }\underset{{\bm \omega}}{\arg\max} \Vert \widehat{{\bm H}}_\text{R} \text{diag}\{{\bm \omega}\} \widehat{{\bm H}}_\text{T} \Vert^2_{\text{F}} = \underset{{{\bm \omega}}}{\arg\max} \Vert {\bm K} {{\bm \omega}} \Vert^2_{2}\\
		\text{s.t.} 
		& \quad \big|[{\bm \omega}]_{[m]}\big| = {1}/{\sqrt{M_{\text{S}}}} , \forall m,
	\end{aligned}
\end{equation}
where ${\bm K} \eqbydef \widehat{{\bm H}}^{\mathsf{T}}_\text{T} \diamond \widehat{{\bm H}}_\text{R}$ \kr{is} obtained by applying Property~1. Note that (\ref{PhiP2}) is non-convex due to the constant modulus constraints. Therefore, we first seek a solution to the following relaxed and convex version of (\ref{PhiP2}) given as
\begin{equation}\label{PhiP5}
	\begin{aligned}
		\mathring{\bm{\omega}}= & \text{ }\underset{\mathring{\bm{\omega}}}{\arg\max} \Vert {\bm K} \mathring{\bm{\omega}} \Vert^2_{2},\quad 
		\text{s.t.} 
		& \Vert\mathring{\bm{\omega}}\Vert_{2} = 1.
	\end{aligned}
\end{equation}

Let ${\bm K} = {\bm U}_{{\bm K}} \bm{\Sigma}_{{\bm K}} {\bm V}^{\mathsf{H}}_{{\bm K}}$ be the SVD of ${\bm K}$. Then, the optimal solution to (\ref{PhiP5}) is given as $\mathring{\bm{\omega}} = [{\bm V}_{{\bm K}}]_{[:,1]}$. To satisfy the constant modulus constraints of (\ref{PhiP2}), we use a simple projection function, where the $m$th entry of ${\bm{\omega}}$ is given as 
\begin{align}\label{fmax}
	[\bm{\omega}^{\text{FroMax-1}}]_{[m]} = \frac{1}{\sqrt{M_{\text{S}}}}   \cdot  \Big(  {	[ \mathring{\bm{\omega}}]_{[m]}   }/{ \big| [ \mathring{\bm{\omega}} ]_{[m]} \big|} \Big).
\end{align}

However, using computer simulations, we have observed that FroMax-1 mainly maximizes the dominant singular value of $\widehat{{\bm H}}_{\text{e}}$, which makes it limited \kr{to single-stream} scenarios. 
%To overcome this issue, we propose in the following the Enhanced FroMax (EFroMax).
%, which increases the rank of the effective communication channel $\widehat{{\bm H}}$.   

\textbf{FroMax-2:} From (\ref{SE2}), we can clearly see that ${\bm{\omega}}$ should be designed so that the singular values $\alpha_{i}$ are maximized. Thus, we propose to modify (\ref{PhiP2}) \kr{as}
\begin{equation}\label{PhiP}
	\begin{aligned}
		{\bm{\omega}} = & \text{ } \underset{{\bm{\omega}} }{\arg\max} \Vert \bm{\Sigma}_s \Vert^2_{\text{F}} = \text{ }\underset{ {\bm{\omega}}  }{\arg\max} \Vert {\bm D} {\bm{\omega}} \Vert^2_{2} \\
		\text{s.t.} 
		& \quad \big|[{\bm{\omega}}]_{[m]}\big| = {1}/{\sqrt{M_{\text{S}}}} , \forall m,
	\end{aligned}
\end{equation}
where ${\bm D} $, due to the diagonal structure of $\bm{\Sigma}_s $, is given as 
\begin{align}\label{Dmat}
	{\bm D} \eqbydef 	\begin{bmatrix}
		[{\bm V}_s]^{\mathsf{T}}_{[:,1]} \widehat{{\bm H}}^{\mathsf{T}}_\text{T} \diamond [{\bm U}_s]^{\mathsf{H}}_{[:,1]} \widehat{{\bm H}}_\text{R} \\
		\vdots \\
		[{\bm V}_s]^{\mathsf{T}}_{[:,N_\text{s}]} \widehat{{\bm H}}^{\mathsf{T}}_\text{T} \diamond [{\bm U}_s]^{\mathsf{H}}_{[:,N_\text{s}]} \widehat{{\bm H}}_\text{R}
	\end{bmatrix} \in \mathbb{C}^{N_\text{s} \times M_\text{S}}.
\end{align}

Similarly to (\ref{PhiP5}), (\ref{PhiP}) can be relaxed to a convex form as 
\begin{equation}\label{PhiP4}
	\begin{aligned}
		\bar{\bm{\omega}} = & \text{ }\underset{\bar{\bm{\omega}}}{\arg\max} \Vert {\bm D} \bar{\bm{\omega}} \Vert^2_{2},\quad 
		\text{s.t.} 
		& \Vert\bar{\bm{\omega}}\Vert_{2} = 1.
	\end{aligned}
\end{equation}

However, differently from (\ref{PhiP5}), we \kr{propose a solution that achieves a higher SE, where} $\bar{\bm{\omega}}$ is obtained by taking the contributions of the dominant $N_\text{s}$ right singular vectors of ${\bm D}$. Specifically, let ${\bm D} = {\bm U}_{{\bm D}} \bm{\Sigma}_{{\bm D}} {\bm V}^{\mathsf{H}}_{{\bm D}}$ be the SVD of ${\bm D}$. Then, \kr{the proposed solution is given as} $\bar{\bm{\omega}} = \frac{[{\bm V}_{{\bm D}}]_{[:,1]} + \dots + [{\bm V}_{{\bm D}}]_{[:,N_\text{s}]}}{\Vert [{\bm V}_{{\bm D}}]_{[:,1]} + \dots + [{\bm V}_{{\bm D}}]_{[:,N_\text{s}]} \Vert_{2}}$. Using $\bar{\bm{\omega}}$, the RIS reflection vector ${\bm{\omega}}$ is obtained as 
\begin{align}\label{efmax}
	[\bm{\omega}^{\text{FroMax-2}}]_{[m]} =   \frac{1}{\sqrt{M_{\text{S}}}}   \cdot  \Big(  {	[ \bar{\bm{\omega}} ]_{[m]}   }/{ \big| [ \bar{\bm{\omega}} ]_{[m]} \big|} \Big), \forall m.
\end{align}

\textbf{Remark 1:} From (\ref{Dmat}), it is clear that the unitary matrices ${\bm U}_s$ and ${\bm V}_s$ are required to construct ${\bm D}$. However, since ${\bm U}_s$ and ${\bm V}_s$ \kr{depend on} $\bm{\omega}$, an iterative two-step algorithm \kr{is} required, where we update ${\bm U}_s$ and ${\bm V}_s$ in one step and $\bm{\omega}$ in the other step. However, we found that if ${\bm U}_s$ and ${\bm V}_s$ are appropriately initialized, then one iteration of such \kr{an} algorithm is sufficient to have a \kr{comparable} SE performance to that obtained by the iterative method of \cite{RISCapacity}. Here, we propose to initialize ${\bm U}_s$ and ${\bm V}_s$ as \kr{follows}. Let $\widehat{{\bm H}}_\text{R} = {\bm U}_{\widehat{{\bm H}}_\text{R}} \bm{\Sigma}_{\widehat{{\bm H}}_\text{R}} {\bm V}^{\mathsf{H}}_{\widehat{{\bm H}}_R}$ and $\widehat{{\bm H}}_\text{T} = {\bm U}_{\widehat{{\bm H}}_\text{T}} \bm{\Sigma}_{\widehat{{\bm H}}_\text{T}} {\bm V}^{\mathsf{H}}_{\widehat{{\bm H}}_\text{T}}$ be the SVD of $\widehat{{\bm H}}_\text{R}$ and $\widehat{{\bm H}}_\text{T}$, respectively. Then, we assume that ${\bm U}_s$ and ${\bm V}_s$ in (\ref{PhiP}) are given as ${\bm U}_s = [{\bm U}_{\widehat{{\bm H}}_\text{R}}]_{[:,1:N_\text{s}]} $ and $ {\bm V}_s = [{\bm V}_{\widehat{{\bm H}}_\text{T}}]_{[:,1:N_\text{s}]}$.

In summary, the proposed beamforming and RIS reflection \kr{coefficient} design method is summarized in Algorithm \ref{Alg1}.

\begin{algorithm}[t]
	\caption{FroMax-based methods for RIS reflection design.}
	\label{Alg1}
	\begin{algorithmic}[1]
		\State{Input: $\widehat{{\bm H}}_{\text{T}}$, $\widehat{{\bm H}}_{\text{R}}$, and $P_{\max}$ }
		\If{FroMax-1 based method}
		\State{Construct ${\bm K}$ as in (\ref{PhiP2}) and get $\mathring{\bm{\omega}}$ from ${\bm V}_{{\bm K}}$}\label{step3}
		\State{Obtain $ \bm{\omega}^{\star}  \leftarrow \bm{\omega}^{\text{FroMax-1}}$ using (\ref{fmax})}
		\ElsIf{FroMax-2 based method}
		\State{Compute ${\bm U}_s = [{\bm U}_{\widehat{{\bm H}}_{\text{R}}}]_{[:,1:N_\text{s}]} $ and $ {\bm V}_s = [{\bm V}_{\widehat{{\bm H}}_{\text{T}}}]_{[:,1:N_\text{s}]}$} \label{step6}
		\State{Construct ${\bm D}$ as in (\ref{Dmat}) and get $\bar{\bm{\omega}}$ from ${\bm V}_{{\bm D}}$} \label{step7}
		\State{Obtain $ \bm{\omega}^{\star}  \leftarrow \bm{\omega}^{\text{FroMax-2}}$ using (\ref{efmax})}
		\EndIf
		\State{\ka{For} given $\bm{\omega}^{\star}$, obtain ${\bm Q}$ and ${\bm P}$ as in (\ref{WF})}\label{step10}
	\end{algorithmic}
\end{algorithm}

\begin{figure*}
	\centering
	\includegraphics[width=\plotsize]{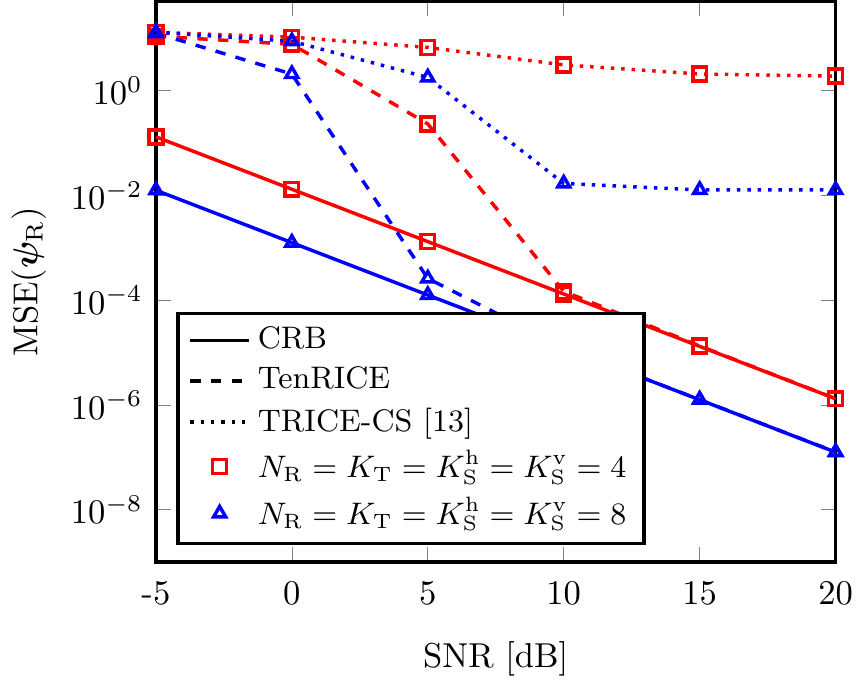}
	\includegraphics[width=\plotsize]{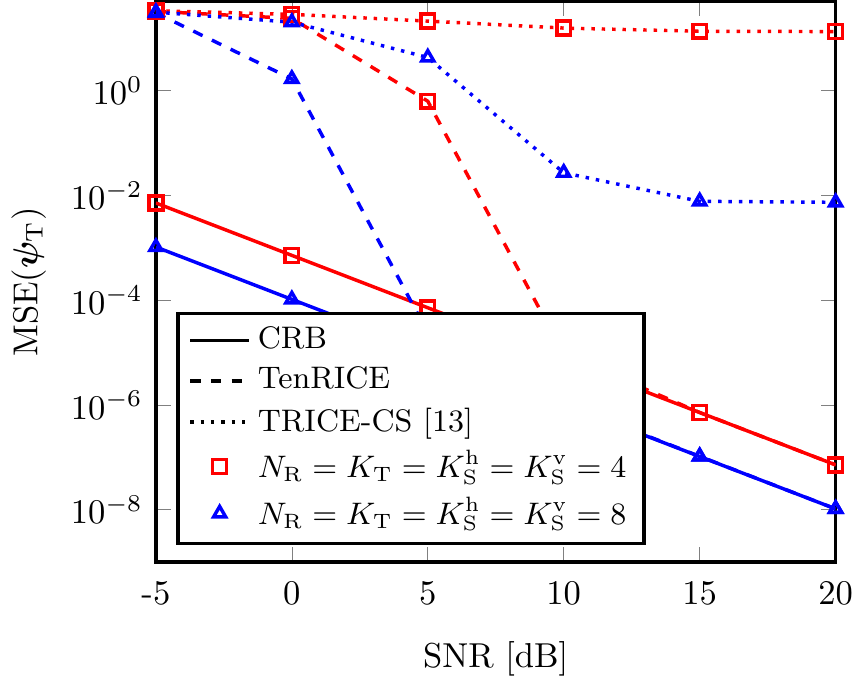}
	\includegraphics[width=\plotsize]{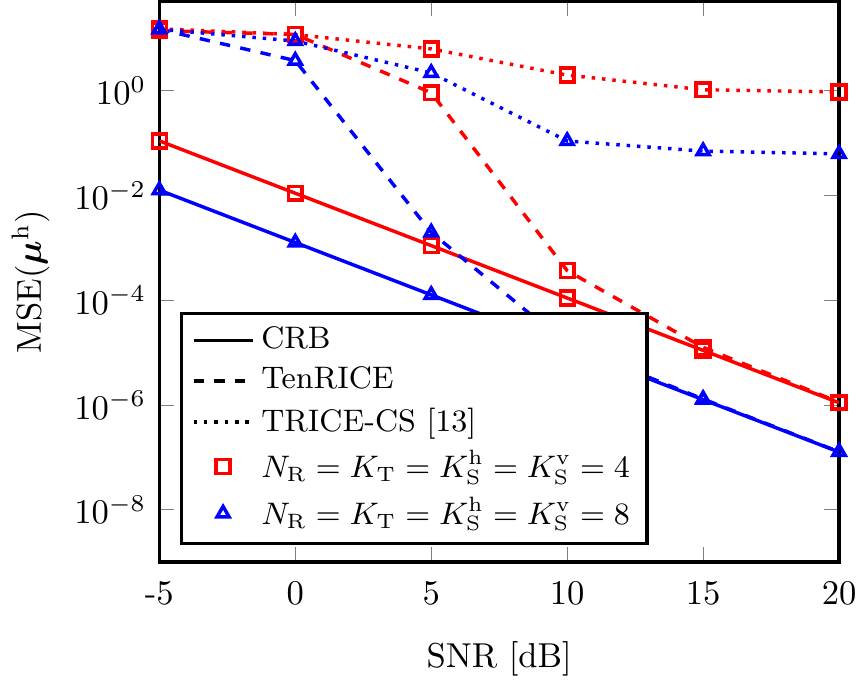}
	\includegraphics[width=\plotsize]{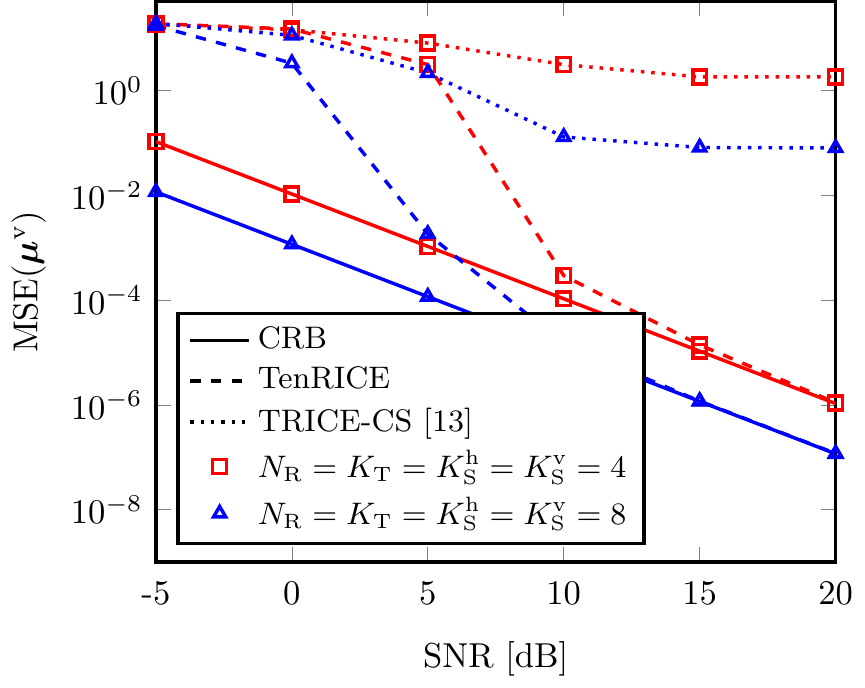}
%	\vspace{-10pt}
	\caption{MSE vs. SNR [$L_\text{T} = L_\text{R} = 2$]. }
	\label{fig2}
	\vspace{-10pt}
\end{figure*}

\begin{figure}
	\centering
	\includegraphics[width=\figsize]{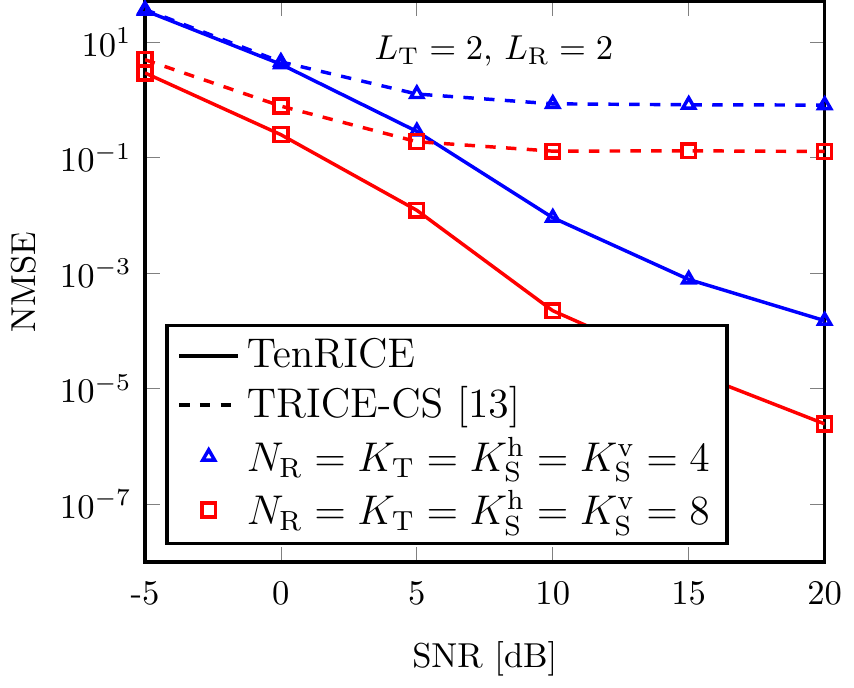}
	\includegraphics[width=\figsize]{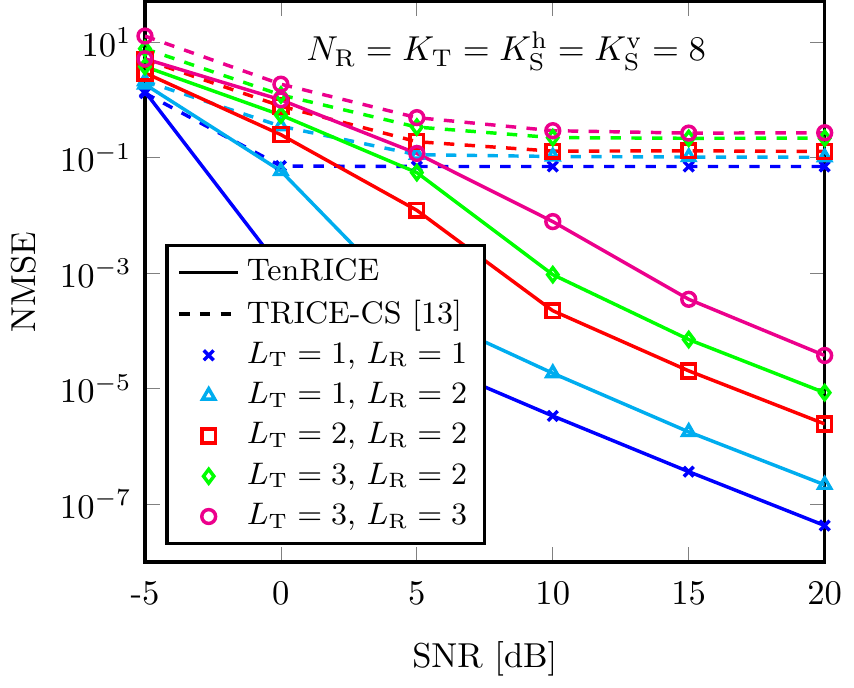}
%	\vspace{-20pt}
	\caption{NMSE vs. SNR.}
	\label{fig3}
%	\vspace{-10pt}
\end{figure}

%\vspace{-5pt}
	
\textbf{Complexity analysis}: \ka{Let the} complexity of calculating the SVD\footnote{Note that the complexity of calculating the SVD of $n\times m$ matrix can be reduced by using the \textit{Power Iteration} method. However, to simplify the analysis, we assume that the SVD is calculated using the bidiagonalization and QR algorithm with a complexity on the order of $\mathcal{O}(nm^2)$.} of a $n\times m$ matrix on the order of $\mathcal{O}(nm^2)$. Then, the complexity of  Algorithm \ref{Alg1} steps \ref{step3}, \ref{step6}, \ref{step7}, and \ref{step10} is \ka{on the} order of $\mathcal{O}(M_\text{R} M_\text{T} M_\text{S}^2)$, $\mathcal{O}(M_\text{R} M_\text{S}^2 + M_\text{S} M_\text{T}^2)$, $\mathcal{O}(N_\text{s} M_\text{S}^2)$, and $\mathcal{O}(M_\text{R} M_\text{T}^2)$ respectively. Accordingly, the complexity of FroMax-1 is \ka{on the} order of $\mathcal{O}\big( M_\text{R} M_\text{T} M_\text{S}^2 +  M_\text{R} M_\text{T}^2 \big)$ and of FroMax-2 is \ka{on the} order of $\mathcal{O}\big( M_\text{R} M_\text{S}^2 + M_\text{S} M_\text{T}^2 + N_\text{s} M_\text{S}^2 +  M_\text{R} M_\text{T}^2 \big)$. In comparison, the complexity of the alternation maximization (AltMax) method of \cite{RISCapacity} is \ka{on the} order of $\mathcal{O}\big( J_{\max} \big( M_\text{S} (3M^3_\text{R} + 2M^2_\text{R} M_\text{T} + M^2_\text{T}) +  M_\text{R} M_\text{T}^2 \big) \big)$, where $J_{\max}$ is the maximum number of iterations.

\section{Numerical Results}
In this section, we show simulation results to evaluate the effectiveness of the proposed methods. 
In all simulation results, we assume that $M_\text{T} = 64, M_\text{R} = 16$, and $ M^{\text{h}}_\text{S} = M^{\text{v}}_\text{S}= 16$, i.e., the RIS has $M_\text{S}= 256$ reflecting elements.

\textbf{Phase 1 - CE}:
In the CE phase, 
%we present simulation results to illustrate the efficiency of the proposed RIS-aided CE method, TenRICE. 
we assume that the training matrices ${\bm W}$, ${\bm F}$, ${\bm \Phi}^{\text{h}}$, and ${\bm \Phi}^{\text{v}}$ in (\ref{eq1}) are randomly generated such that the $(i,j)$th entry of ${\bm W}$ is given as $[{\bm W}]_{[i,j]} = \frac{1}{\sqrt{M}_{\text{R}}}  e^{j \varphi_{i,j}}, \varphi_{i,j} \in [0,2\pi]$ , where ${\bm F}$, ${\bm \Phi}^{\text{h}}$, and ${\bm \Phi}^{\text{v}}$ are similarly generated. We show results in terms of the mean-squared error (MSE) of $\bm{\psi}_{\text{R}}$ defined as $\text{MSE}(\bm{\psi}_{\text{R}}) = \mathbb{E}\big\{ \Vert \bm{\psi}_{\text{R}} - \hat{\bm{\psi}}_{\text{R}} \Vert^{2}_{2} \big\}$, where $\text{MSE}(\bm{\psi}_{\text{T}})$, $\text{MSE}(\bm{\mu}^{\text{h}} )$, and $\text{MSE}(\bm{\mu}^{\text{v}} )$ are similarly defined, and the normalized MSE (NMSE) of the cascaded channel \kr{is} defined as \ka{$\text{NMSE} = \mathbb{E} \big\{{\Vert {\bm H}_{\text{c}} - \widehat{{\bm H}}_{\text{c}} \Vert^{2}_{\text{F}} \big\} }/\mathbb{E} \big\{{\Vert {\bm H}_{\text{c}}\Vert^{2}_{\text{F}}} \big\}$.} We define \ka{the} signal-to-noise ratio (SNR) as \ka{$\text{SNR} = \mathbb{E} \big\{{\Vert \bm{\mathcal{Y}} - \bm{\mathcal{Z}}\Vert^{2}_{\text{F}}  }\big\}/\mathbb{E}\big\{{\Vert \bm{\mathcal{Z}}\Vert^{2}_{\text{F}}} \big\}$. For comparison, we include simulation results of the two-stage TRICE-CS framework \cite{ardah2020trice}, where the estimation is {performed} using the classical OMP technique \cite{OMPComp} assuming a 2D dictionary of $ 128 \times 128$ grid points in both stages.}

\begin{figure}[t]
	\centering
	\includegraphics[width=\figsize]{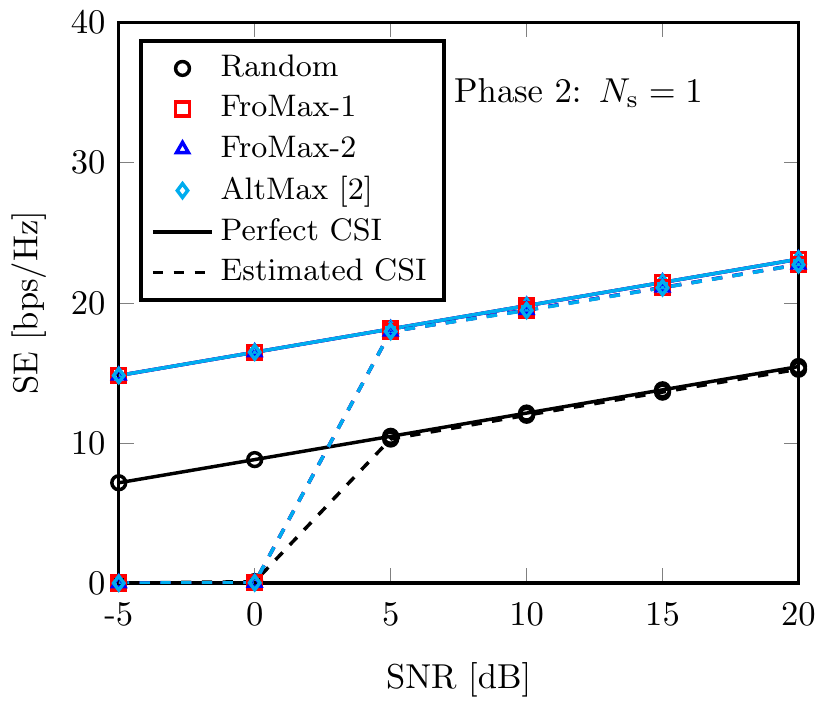}
	\includegraphics[width=\figsize]{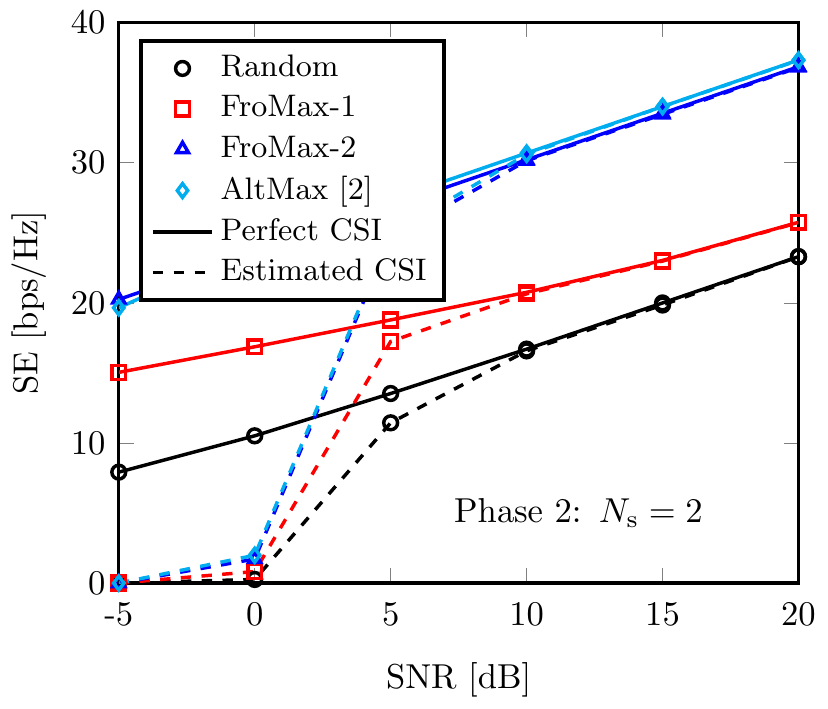}
%		\vspace{-20pt}
	\caption{SE vs. SNR. [$L_\text{T} = L_\text{R} = 2$. Phase 1: $K_\text{R} = K_\text{T} = K^\text{h}_\text{S} = K^\text{v}_\text{S} = 8$]}
	\label{fig4}
	\vspace{-15pt}
\end{figure}
\begin{figure}[t]
	\centering
	\includegraphics[width=\figsize]{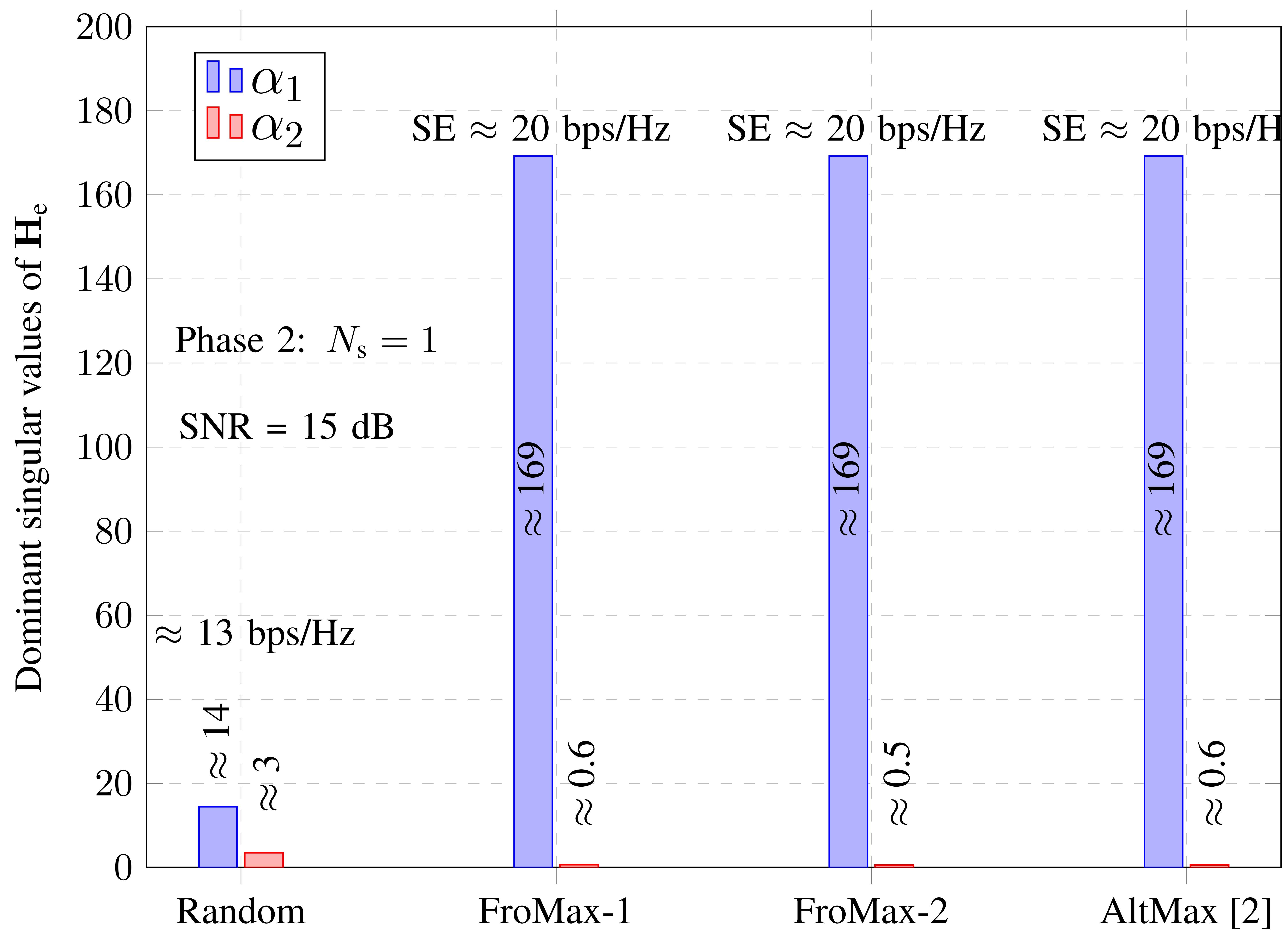}
	\includegraphics[width=\figsize]{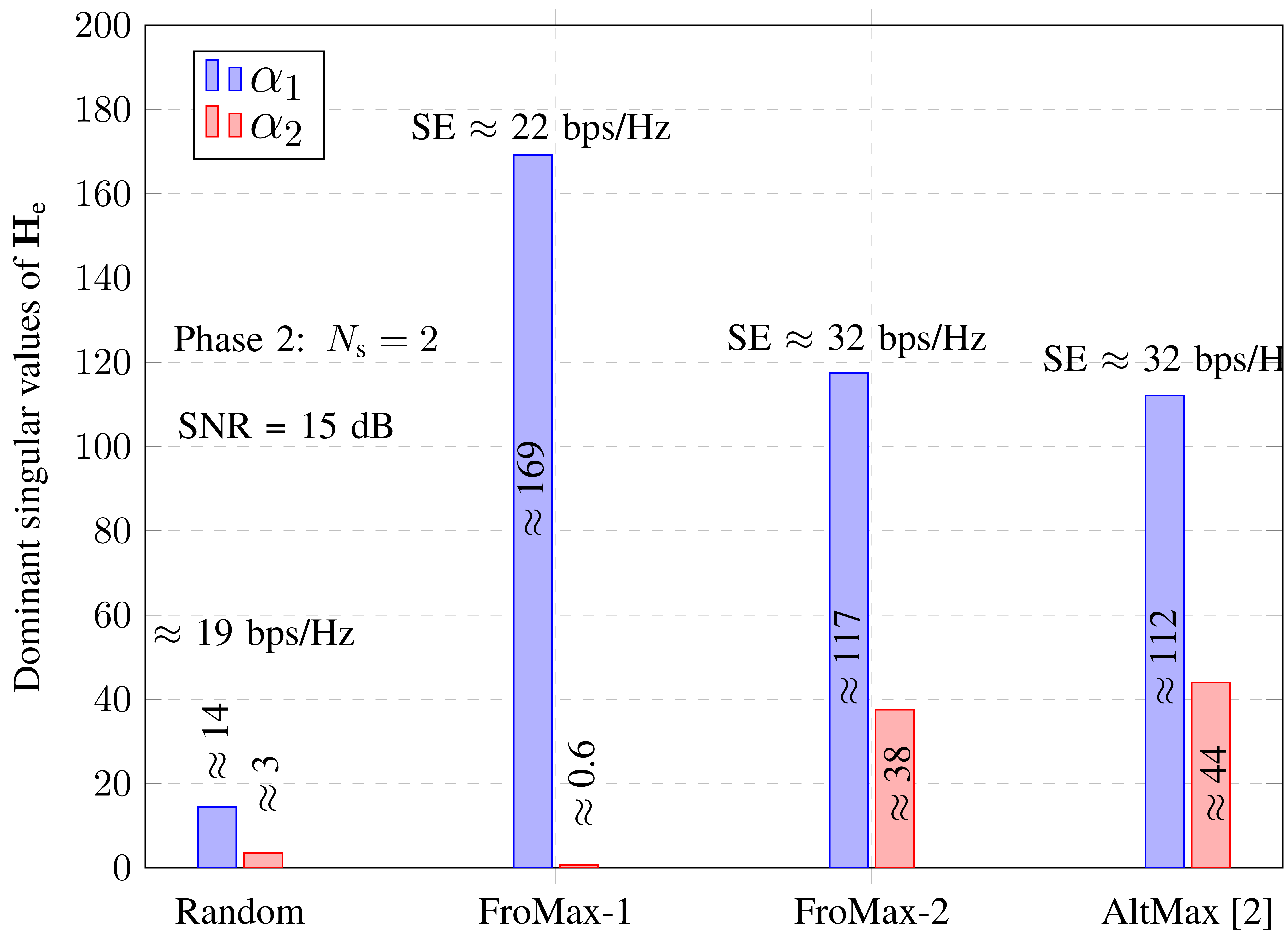}
%		\vspace{-20pt}
	\caption{Dominant singular values of a perfect effective channel ${\bm H}_{\text{e}} = {\bm H}_{\text{R}} \text{diag}\{{\bm \omega}\} {\bm H}_{\text{T}}$ for one channel realization [$L_\text{T} = L_\text{R} = 2$].}
	\label{fig5}
%	\vspace{-10pt}
\end{figure}

Figs. \ref{fig2} and \ref{fig3} show \kr{the} MSE versus \kr{the} SNR and \kr{the} NMSE versus \kr{the} SNR results, respectively, averaged over 1,000 channel realizations. From Fig. \ref{fig2}, we can see that TenRICE provides \kr{more} accurate parameter estimates, compared to TRICE-CS, approaching the CRB\footnote{The CRB derivation to our 4-way CP tensor is a straightforward extension of the CRB derivation in \cite{LowRankTen} for a 3-way CP tensor. Therefore, it has been omitted here due to brevity.} as \ka{the} SNR increases. The main reason is that TenRICE not only exploits the low-rank nature of mmWave channels, but also the tensor structure of the received signals when estimating the channel parameters. Moreover, TenRICE employs a high-resolution parameter recovery method in NOMP, while TRICE-CS suffers from quantization errors, due to the on-grid assumption. These advantages lead to more accurate \kr{channel estimates}, as can be seen from Fig.~\ref{fig3}, with less training overhead and lower complexity. %\ka{Here, we note that the ALS steps of TenRICE method requires a small number of iterations $I_{\max}$ to converge, e.g., $I_{\max} = 10$ for SNR = 5 dB and $I_{\max} = 30$ for SNR = 20 dB.}  

% \subsection{Performance evaluation of FroMax}\label{subsec2}
\textbf{Phase 2 - DT}:
Next, we show simulation results to illustrate the efficiency of the proposed RIS reflection design method, FroMax. For comparison, we include results when the RIS reflection \kr{coefficient} vector ${\bm \omega}$ is designed according to the alternating maximization method in \cite{RISCapacity}, termed AltMax, and Random, where the entries of ${\bm \omega}$ are randomly generated such that the $m$th entry is given as $[{\bm \omega}]_{[m]} = \frac{1}{\sqrt{M_{\text{S}}}} e^{j\omega_m}, \omega_m\in [0,2\pi]$. We define \ka{the} SNR as \ka{$\text{SNR} = P_{\max}/\sigma^2$.}

Fig. \ref{fig4} shows SE versus SNR results, averaged over 1,000 channel realizations. Clearly, we can see that FroMax-1 has an equal performance to that of FroMax-2 and AltMax when $N_\text{s} = 1$. However, FroMax-1 experiences a performance loss  when $N_\text{s} = 2$, since it mainly maximizes the \kr{dominant} singular value, as it can be seen from Fig.~\ref{fig5}. Differently, \kr{the} AltMax and FroMax-2 methods optimize the dominant $N_\text{s}$ singular values of the effective channel \kr{such} that it maximizes the system SE. Note that, in the low SNR regime, i.e., below 5 dB, all the simulated methods experience a very low SE performance, due to the CE errors. Therefore, a preprocessing \textit{denoising} step will be required to improve the CE accuracy, which we leave for future work.

%	\vspace{-5pt}
\section{Conclusions}
%	\vspace{-5pt}
In this work, we have considered the channel estimation and the RIS reflection \kr{coefficient} design problems in point-to-point RIS-aided mmWave MIMO communication systems. We have proposed a CP tensor-based channel estimation method termed TenRICE, which estimates the transmitter to RIS and the RIS to receiver channels separately, up to a trivial scaling factor. We have shown that by jointly exploiting the low-rank nature of mmWave channels and the tensor structure of \kr{the} received signals, not only the estimation accuracy can be improved, but also the training overhead and the complexity can be reduced. \ka{ The proposed} non-iterative RIS reflection design method based on a Frobenius-norm maximization (FroMax) design strategy has a comparable performance to a benchmark method but with significantly lower complexity.

% \newpage	
\bibliographystyle{IEEEtran}
\bibliography{refs}	

\begin{thebibliography}{10}
\providecommand{\url}[1]{#1}
\csname url@samestyle\endcsname
\providecommand{\newblock}{\relax}
\providecommand{\bibinfo}[2]{#2}
\providecommand{\BIBentrySTDinterwordspacing}{\spaceskip=0pt\relax}
\providecommand{\BIBentryALTinterwordstretchfactor}{4}
\providecommand{\BIBentryALTinterwordspacing}{\spaceskip=\fontdimen2\font plus
\BIBentryALTinterwordstretchfactor\fontdimen3\font minus
  \fontdimen4\font\relax}
\providecommand{\BIBforeignlanguage}[2]{{%
\expandafter\ifx\csname l@#1\endcsname\relax
\typeout{** WARNING: IEEEtran.bst: No hyphenation pattern has been}%
\typeout{** loaded for the language `#1'. Using the pattern for}%
\typeout{** the default language instead.}%
\else
\language=\csname l@#1\endcsname
\fi
#2}}
\providecommand{\BIBdecl}{\relax}
\BIBdecl

\bibitem{irs}
M.~Di~Renzo, A.~Zappone, M.~Debbah, M.-S. Alouini, C.~Yuen, J.~de~Rosny, and
  S.~Tretyakov, ``Smart radio environments empowered by reconfigurable
  intelligent surfaces: How it works, state of research, and the road ahead,''
  \emph{IEEE J. Sel. Areas Commun.}, vol.~38, no.~11, pp. 2450--2525, 2020.

\bibitem{RISCapacity}
S.~Zhang and R.~Zhang, ``Capacity characterization for intelligent reflecting
  surface aided {MIMO} communication,'' \emph{IEEE J. Sel. Areas Commun.},
  vol.~38, no.~8, pp. 1823--1838, Aug. 2020.

\bibitem{BF1}
Q.~{Wu} and R.~{Zhang}, ``Intelligent reflecting surface enhanced wireless
  network via joint active and passive beamforming,'' \emph{IEEE Trans.
  Wireless Commun.}, vol.~18, no.~11, pp. 5394--5409, 2019.

\bibitem{RISsec}
L.~Dong and H.-M. Wang, ``Enhancing secure {MIMO} transmission via intelligent
  reflecting surface,'' \emph{IEEE Trans. Wireless Commun.}, vol.~19, no.~11,
  pp. 7543--7556, Nov. 2020.

\bibitem{IRSSINR}
Q.-U.-A. Nadeem, A.~Kammoun, A.~Chaaban, M.~Debbah, and M.-S. Alouini,
  ``Asymptotic max-min {SINR} analysis of reconfigurable intelligent surface
  assisted {MISO} systems,'' \emph{IEEE Trans. Wireless Commun.}, vol.~19,
  no.~12, pp. 7748--7764, Dec. 2020.

\bibitem{LSOnOff}
D.~{Mishra} and H.~{Johansson}, ``Channel estimation and low-complexity
  beamforming design for passive intelligent surface assisted {MISO} wireless
  energy transfer,'' in \emph{Proc. IEEE International Conference on Acoustics,
  Speech and Signal Processing (ICASSP)}, 2019, pp. 4659--4663.

\bibitem{MVDR}
T.~L. {Jensen} and E.~{De Carvalho}, ``An optimal channel estimation scheme for
  intelligent reflecting surfaces based on a minimum variance unbiased
  estimator,'' in \emph{Proc. IEEE International Conference on Acoustics,
  Speech and Signal Processing (ICASSP)}, 2020, pp. 5000--5004.

\bibitem{MMSE}
Q.~{Nadeem}, H.~{Alwazani}, A.~{Kammoun}, A.~{Chaaban}, M.~{Debbah}, and
  M.~{Alouini}, ``Intelligent reflecting surface-assisted multi-user {MISO}
  communication: Channel estimation and beamforming design,'' \emph{IEEE Open
  J. Commun. Soc.}, vol.~1, pp. 661--680, 2020.

\bibitem{RISUL}
J.~{Zhang}, C.~{Qi}, P.~{Li}, and P.~{Lu}, ``Channel estimation for
  reconfigurable intelligent surface aided massive {MIMO} system,'' in
  \emph{Proc. IEEE 21st International Workshop on Signal Processing Advances in
  Wireless Communications (SPAWC)}, May 2020, pp. 1--5.

\bibitem{CS}
D.~L. {Donoho}, ``Compressed sensing,'' \emph{IEEE Trans. Inf. Theory},
  vol.~52, no.~4, pp. 1289--1306, Apr. 2006.

\bibitem{ardah_icassp2020}
K.~{Ardah}, B.~{Sokal}, A.~L.~F. {de Almeida}, and M.~{Haardt}, ``Compressed
  sensing based channel estimation and open-loop training design for hybrid
  analog-digital massive {MIMO} systems,'' in \emph{Proc. IEEE International
  Conference on Acoustics, Speech and Signal Processing (ICASSP)}, May. 2020,
  pp. 4597--4601.

\bibitem{ardah_icassp19}
K.~{Ardah}, A.~L.~F. {de Almeida}, and M.~{Haardt}, ``A gridless {CS} approach
  for channel estimation in hybrid massive {MIMO} systems,'' in \emph{Proc.
  IEEE International Conference on Acoustics, Speech and Signal Processing
  (ICASSP)}, May 2019, pp. 4160--4164.

\bibitem{ardah2020trice}
K.~Ardah, S.~Gherekhloo, A.~L.~F. de~Almeida, and M.~Haardt, ``{TRICE}: A
  channel estimation framework for {RIS}-aided millimeter-wave {MIMO}
  systems,'' \emph{IEEE Signal Process. Lett.}, vol.~28, pp. 513--517, Feb.
  2021.

\bibitem{CSGrid}
P.~{Wang}, J.~{Fang}, H.~{Duan}, and H.~{Li}, ``Compressed channel estimation
  for intelligent reflecting surface-assisted millimeter wave systems,''
  \emph{IEEE Signal Process. Lett.}, vol.~27, pp. 905--909, 2020.

\bibitem{MatrixCom}
Z.~{He} and X.~{Yuan}, ``Cascaded channel estimation for large intelligent
  metasurface assisted massive {MIMO},'' \emph{IEEE Wireless Commun. Lett.},
  vol.~9, no.~2, pp. 210--214, Feb. 2020.

\bibitem{Andre2014}
G.~{Favier} and A.~L. {de Almeida}, ``{Overview of constrained {PARAFAC}
  models},'' \emph{EURASIP Journal on Applied Signal Processing}, vol. 2014, p.
  142, Dec. 2014.

\bibitem{andre}
G.~T. {de Araújo} and A.~L.~F. {de Almeida}, ``{PARAFAC}-based channel
  estimation for intelligent reflective surface assisted {MIMO} system,'' in
  \emph{Proc. IEEE 11th Sensor Array and Multichannel Signal Processing
  Workshop (SAM)}, 2020, pp. 1--5.

\bibitem{Rappaport}
T.~S. {Rappaport}, Y.~{Xing}, G.~R. {MacCartney}, A.~F. {Molisch},
  E.~{Mellios}, and J.~{Zhang}, ``Overview of millimeter wave communications
  for fifth-generation ({5G}) wireless networks—with a focus on propagation
  models,'' \emph{IEEE Transactions on Antennas and Propagation}, vol.~65,
  no.~12, pp. 6213--6230, 2017.

\bibitem{StESBRIT}
J.~{Zhang} and M.~{Haardt}, ``Channel estimation and training design for hybrid
  multi-carrier mmwave massive {MIMO} systems: The beamspace {ESPRIT}
  approach,'' in \emph{Proc. 25th European Signal Processing Conference
  (EUSIPCO)}, 2017, pp. 385--389.

\bibitem{Andre2008}
A.~L.~F. de~Almeida, G.~Favier, and J.~C.~M. Mota, ``A constrained factor
  decomposition with application to {MIMO} antenna systems,'' \emph{IEEE Trans.
  Signal Process.}, vol.~56, no.~6, pp. 2429--2442, 2008.

\bibitem{Lathauwer}
L.~De~Lathauwer, ``A link between the canonical decomposition in multilinear
  algebra and simultaneous matrix diagonalization,'' \emph{SIAM Journal on
  Matrix Analysis and Applications}, vol.~28, no.~3, pp. 642--666, Apr. 2006.

\bibitem{ardah2019wsa}
K.~{Ardah}, A.~L.~F. de~{Almeida}, and M.~{Haardt}, ``Low-complexity millimeter
  wave {CSI} estimation in {MIMO-OFDM} hybrid beamforming systems,'' in
  \emph{Proc. 23rd International ITG Workshop on Smart Antennas (WSA)}, Apr.
  2019, pp. 1--5.

\bibitem{ROEMER20132722}
F.~Roemer and M.~Haardt, ``A semi-algebraic framework for approximate {CP}
  decompositions via simultaneous matrix diagonalizations {(SECSI)},''
  \emph{Signal Processing}, vol.~93, no.~9, pp. 2722 -- 2738, 2013.

\bibitem{andre_tensors}
P.~Comon, X.~Luciani, and A.~L.~F. de~Almeida, ``Tensor decompositions,
  alternating least squares and other tales,'' \emph{Journal of Chemometrics},
  vol.~23, no. 7‐8, pp. 393--405, 2009.

\bibitem{NOMP}
B.~{Mamandipoor}, D.~{Ramasamy}, and U.~{Madhow}, ``Newtonized orthogonal
  matching pursuit: Frequency estimation over the continuum,'' \emph{IEEE
  Trans. Signal Process.}, vol.~64, no.~19, pp. 5066--5081, Oct. 2016.

\bibitem{LowRankTen}
Z.~{Zhou}, J.~{Fang}, L.~{Yang}, H.~{Li}, Z.~{Chen}, and R.~S. {Blum},
  ``Low-rank tensor decomposition-aided channel estimation for millimeter wave
  {MIMO-OFDM} systems,'' \emph{IEEE J. Sel. Areas Commun.}, vol.~35, no.~7, pp.
  1524--1538, Jul. 2017.

\bibitem{Kolda}
T.~G. Kolda and B.~W. Bader, ``Tensor decompositions and applications,''
  \emph{SIAM Review}, vol.~51, no.~3, pp. 455--500, Sept. 2009.

\bibitem{CCPD}
A.~L.~F. de~Almeida, G.~Favier, and J.~C.~M. Mota, ``Constrained tensor
  modeling approach to blind multiple-antenna {CDMA} schemes,'' \emph{IEEE
  Trans. Signal Process.}, vol.~56, no.~6, pp. 2417--2428, Jun. 2008.

\bibitem{Andre2010}
A.~Stegeman and A.~L.~F. de~Almeida, ``Uniqueness conditions for constrained
  three-way factor decompositions with linearly dependent loadings,''
  \emph{SIAM Journal on Matrix Analysis and Applications}, vol.~31, no.~3, pp.
  1469--1490, 2010.

\bibitem{OMPComp}
B.~L. {Sturm} and M.~G. {Christensen}, ``Comparison of orthogonal matching
  pursuit implementations,'' in \emph{Proc. of the 20th European Signal
  Processing Conference (EUSIPCO)}, Aug. 2012, pp. 220--224.

\bibitem{heath_overview}
R.~W. {Heath}, N.~{González-Prelcic}, S.~{Rangan}, W.~{Roh}, and A.~M.
  {Sayeed}, ``An overview of signal processing techniques for millimeter wave
  {MIMO} systems,'' \emph{IEEE J. Sel. Topics Signal Process.}, vol.~10, no.~3,
  pp. 436--453, 2016.

\bibitem{sepideh}
S.~{Gherekhloo}, K.~{Ardah}, and M.~{Haardt}, ``Hybrid beamforming design for
  downlink {MU-MIMO-OFDM} millimeter-wave systems,'' in \emph{Proc. IEEE 11th
  Sensor Array and Multichannel Signal Processing Workshop (SAM)}, Jun. 2020,
  pp. 1--5.

\bibitem{ardah_unify}
K.~{Ardah}, G.~{Fodor}, Y.~C.~B. {Silva}, W.~C. {Freitas}, and F.~R.~P.
  {Cavalcanti}, ``A unifying design of hybrid beamforming architectures
  employing phase shifters or switches,'' \emph{IEEE Trans. Veh. Technol.},
  vol.~67, no.~11, pp. 11\,243--11\,247, Nov. 2018.

\bibitem{ardah_tvt}
K.~{Ardah}, G.~{Fodor}, Y.~C.~B. {Silva}, W.~C. {Freitas}, and A.~L.~F. {de
  Almeida}, ``Hybrid analog-digital beamforming design for {SE} and {EE}
  maximization in massive {MIMO} networks,'' \emph{IEEE Trans. Veh. Technol.},
  vol.~69, no.~1, pp. 377--389, Jan. 2020.

\bibitem{Palomar2005}
D.~{Palomar} and J.~Fonollosa, ``Practical algorithms for a family of
  waterfilling solutions,'' \emph{IEEE Trans. Signal Process.}, vol.~53, no.~2,
  pp. 686--695, 2005.

\end{thebibliography}

%\vfill\pagebreak
%\bibliographystyle{IEEEbib}
%\bibliography{strings,refs}

%\section{REFERENCES}
%\label{sec:refs}


\end{document}